\documentclass{aa}
\usepackage{graphicx}
\usepackage{natbib}
\usepackage{epstopdf}
\bibpunct{(}{)}{;}{a}{}{,}
\usepackage{txfonts}

\begin{document}
\title{Resolved photometry of extragalactic young massive star clusters\thanks{Based on observations made with the NASA/ESA Hubble Space Telescope, obtained from the data archive at the Space Telescope Institute. STScI is operated by the association of Universities for Research in Astronomy, Inc. under the NASA contract  NAS 5-26555}
}

\author{S. S. Larsen \inst{1} \and
  S. E. de Mink\thanks{Hubble Fellow} \inst{2} \and
  J. J. Eldridge \inst{3} \and
  N. Langer \inst{4} \and
  N. Bastian \inst{5} \and
  A. Seth \inst{6} \and
  L. J. Smith \inst{2}$^,$\inst{9} \and
  J. Brodie \inst{7} \and
  Yu. N. Efremov \inst{8}
}
\institute{Astronomical Institute, University of Utrecht, Princetonplein 5,
  NL-3584 CC, Utrecht, The Netherlands
\and
  Space Telescope Science Institute, 3700 San Martin Drive, Baltimore, MD 21218, USA
\and
  Institute of Astronomy, The Observatories, University of Cambridge, Madingley Road, Cambridge, CB3 0HA, United Kingdom
\and
  Argelander-Institut f{\"u}r Astronomie, Bonn University, Auf dem H{\"u}gel 71, 53121 Bonn, Germany
\and
  Excellence Cluster Universe, Boltzmannstr. 2, 85748 Garching, Germany
\and        
  Harvard-Smithsonian Center for Astrophysics, 60 Garden Street, Cambridge, MA 02138, USA
\and
  UCO/Lick Observatory, University of California, Santa Cruz, CA 95064, USA
\and
  Sternberg Astronomical Institute, MSU, Moscow 119992, Universitetsky pr. 13, Russia
\and
   European Space Agency, Research and Scientific Support Department, Baltimore, MD 21218, USA 
}

\offprints{S.\ S.\ Larsen, \email{s.s.larsen@astro-uu.nl}}

\date{A\&A, accepted}

\abstract
{}
{We present colour-magnitude diagrams (CMDs) of young massive star clusters in several galaxies located well beyond the Local Group. The richness of these clusters allows us to obtain large samples of post-main sequence stars and test how well the observed CMDs are reproduced by canonical stellar isochrones.}
{We use imaging of seven clusters in the galaxies \object{NGC~1313}, \object{NGC~1569}, \object{NGC~1705}, \object{NGC~5236} and \object{NGC~7793} obtained with the Advanced Camera for Surveys on board the \emph{Hubble Space Telescope} and carry out PSF-fitting photometry of individual stars in the clusters. The clusters have ages in the range $\sim(5-50)\times10^6$ years and masses of $\sim10^5 M_\odot$--$10^6 M_\odot$. Although crowding prevents us from obtaining photometry in the inner regions of the clusters, we are still able to measure up to 30--100 supergiant stars in each of the richest clusters. The resulting CMDs and luminosity functions are compared with photometry of artificially generated clusters, designed to reproduce the photometric errors and completeness as realistically as possible. 
}
{In agreement with previous studies, our CMDs show no clear gap between the H-burning main sequence and the He-burning supergiant stars, contrary to predictions by common stellar isochrones. In general, the isochrones also fail to match the observed number ratios of red-to-blue supergiant stars, although the difficulty of separating blue supergiants from the main sequence complicates this comparison. In several cases we observe a large spread (1--2 mag) in the luminosities of the supergiant stars that cannot be accounted for by observational errors. We find that this spread can be reproduced by including an age spread of $\sim(10-30)\times10^6$ years in the models. However, age spreads cannot fully account for the observed morphology of the CMDs and other processes, such as the evolution of interacting binary stars, may also play a role.
}
{Colour-magnitude diagrams can be successfully obtained for massive star clusters out to distances of at least 4--5 Mpc. Comparing such CMDs with models based on canonical isochrones we find several areas of disagreement. One interesting possibility is that an age spread of up to $\sim30$ Myr may be present in some clusters. The data presented here may provide useful constraints on models for single and/or binary stellar evolution.
}


\titlerunning{Colour-magnitude diagrams of YMCs}
\maketitle

\section{Introduction}

Stars profoundly affect the properties of the galaxies they inhabit. They are the primary sites of nucleosynthesis and drivers of chemical evolution and they contribute to the energy budget of the interstellar medium through feedback processes such as supernovae and stellar winds. In most galaxies they also deliver most of the ultraviolet, optical and near-infrared light. In order to correctly interpret observations of galaxies and constrain their past evolution (e.g.\ in terms of star formation histories and chemical evolution) it is therefore essential to strive towards the best possible understanding of the properties of individual stars. This is true not only for integrated light measurements, but also for modelling of colour-magnitude diagrams (CMDs) of resolved stellar populations \citep{Gallart2005}.

Although stars spend most of their lifetimes in the core-hydrogen burning phase on the main sequence (MS), post-MS stars provide a significant fraction of the bolometric luminosity of a stellar population at all except the very youngest (few Myr) ages \citep[e.g.][]{Renzini1986}. Intermediate- and high mass post-MS stars contribute significantly to the energy output of young stellar populations, and the modelling of such stars is particularly uncertain. One example is the relative numbers of blue and red core-He burning stars, which is very sensitive to assumptions about input parameters such as mass loss and rotation \citep{Langer1995,Maeder2000}. Models are not yet capable of reproducing trends in the red/blue ratio e.g.\ as a function of metallicity seen in star clusters \citep{Eggenberger2002}, although somewhat better agreement is found when comparing trends in  this ratio versus \emph{age} \citep{Dohm-Palmer1997,Dohm-Palmer2002}. 
Another issue concerns the extent of the ``blue loops'' in the core-He burning phase, which depends on the detailed structure and chemical profile of a star. These, in turn, are affected e.g. by the treatment of convective overshooting \citep{Matraka1982,Ritossa1996}, with the net effect that more overshooting tends to lead to less extended (redder) blue loops.
 For standard assumptions, models tend to underpredict the extent of the blue loops. While most stellar evolution models predict a gap in the colour-magnitude diagram (CMD) between the H and He burning stars 
\citep[the ``Blue Hertzsprung Gap'' or BHG;][]{Chiosi1998,Salasnich1999}, the post-MS radius evolution is very uncertain for massive stars and star cluster CMDs often show significant numbers of stars in the BHG \citep[e.g.][]{Fischer1993}. 
The nature of these stars is poorly understood and it is unclear whether they are even core H or He-burning \citep{Vink2010}, although in some cases their CNO abundances suggest that they have undergone a first dredge-up as red supergiants \citep{Salasnich1999}. 
Some models do predict more extended blue loops, but this requires that a very large fraction of the H-rich envelope is lost in the red supergiant stage \citep{Hirschi2004,Salasnich1999}. This may not be compatible with the fact that SN 1987A still had a massive H-rich envelope when it exploded as a blue supergiant \citep{Woosley1988}.
Complicating matters further a large fraction of massive stars are members of binary systems \citep{Sana2008,Mason2009}, many of which may interact or even merge during a phase of common envelope evolution \citep{Taam2000,Eldridge2008,Sansom2009}.

Historically, star clusters have long served as important testbeds for stellar models.
A prime reason for this has been the assumption that all stars in a cluster are very nearly coeval, so that a cluster CMD can be directly compared against a model isochrone. Furthermore, under this assumption all post-MS stars in a given cluster will have very similar (initial) masses, so that the step of interpolation in the tracks to produce isochrones can often be eliminated for comparison with observed CMDs or calculation of integrated properties \citep{Maraston1998}. 
Models of low-mass stars in particular have been extensively tested by comparison with globular cluster CMDs \citep{Renzini1988,Palmieri2002,Salaris2007}, in some cases providing evidence of multiple stellar populations \citep{Piotto2009}.  Similarly, high precision photometry has shown that the CMDs of several intermediate-age ($\sim1-2$ Gyr) clusters in the Magellanic Clouds are not well fitted by standard isochrones, but instead show a significant spread in the main sequence turn-off magnitudes and luminosities of red clump stars. One possible explanation for these features is a significant age spread in these clusters  \citep{Mackey2008,Milone2009}, although other effects (such as stellar rotation) may also play a role (\citealt{Bastian2009}; but see also \citealt{Girardi2011}).
However, interpretation of cluster CMDs remains easier than the complex mix of stellar populations that make up the field, and clusters remain attractive targets for comparison with stellar models.
In fact, such a comparison may yield important information on possible age spreads within clusters, in addition to providing constraints on the stellar models.

\begin{figure}
\centering
\includegraphics[width=85mm]{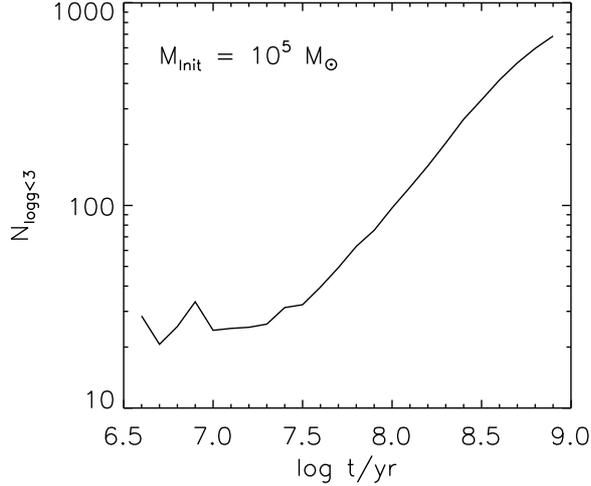}
\caption{\label{fig:npms}Number of stars with $\log g < 3.0$ for an initial total stellar mass of $10^5 M_\odot$. 
 }
\end{figure}

One difficulty is that a typical low-mass young (open) cluster contains no or only a few post-MS stars. 
As an illustration, Fig.~\ref{fig:npms} shows the number of stars with $\log g < 3.0$ vs.\ age, predicted for a single-age stellar population with an initial mass of $10^5 M_\odot$. We have assumed a Salpeter IMF over the mass range $0.15 < M/M_\odot < 100$ and used Padua isochrones \citep{Bertelli2009} to obtain $\log g$ values versus age and (initial) stellar mass. The $\log g$ limit adopted in this figure comfortably includes all local supergiants \citep{Lyubimkov2010}. The figure shows that, for ages less than about 30 Myr, only 20--30 such stars are expected per $10^5 M_\odot$, with an increase at older ages. The figure is consistent with observations of post-MS stars in massive star clusters, e.g. of the rich \object{LMC} cluster \object{NGC~1866} which has 100 post-MS stars, an age of about 160 Myr and a mass of about $1.3\times10^5 M_\odot$ \citep{Fischer1992,Barmina2002}, or the ``\object{Red Supergiant Cluster 2}'' (age $\sim10$ Myr, initial mass $\sim$ 40\,000 $M_\odot$) which hosts 26 red supergiant stars \citep{Davies2007}. It is thus clear that cluster masses of at least several times $10^4 M_\odot$ are required in order to get significant samples of massive post-MS stars. Such clusters are rare in the Milky Way and even in the Local Group, making it difficult to obtain samples that cover a range in age (or, equivalently, main sequence turn-off mass), metallicity, etc.

If the search volume is extended beyond the boundaries of the Local Group several galaxies hosting young clusters with $M\ga10^5 M_\odot$ become available. The trade-off is that photometry of individual stars becomes extremely challenging due to severe crowding, even if imaging from the Hubble Space Telescope (HST) is used. One of the closest examples of a galaxy hosting such massive, young star clusters is \object{NGC~1569} at $\sim3.4$ Mpc \citep{Grocholski2008}. Using imaging obtained with the High Resolution Channel (HRC) of the Advanced Camera for Surveys (ACS) on board HST, \citet{Larsen2008a} constructed a CMD for the cluster \object{NGC~1569-B} and identified about 60 red supergiant stars (RSGs). This demonstrates that photometry can indeed be obtained for individual stars in clusters well beyond the Local Group, and motivated us to search for additional candidates suitable for such analysis.
Several other galaxies located within distances of a few Mpc have ACS imaging, obtained as part of our own and other programmes. We have inspected these images and looked for clusters that appeared sufficiently well resolved into individual stars that photometry would be feasible. In this paper we present the resulting CMDs and compare with theoretical isochrones. Our aim is to quantify whether or not the observations can be well approximated by single isochrones.

\section{Data reduction and analysis}

\subsection{Selection of the sample}

\begin{table*}
\begin{minipage}[t]{15cm}
\caption{Data for the clusters analysed in this paper. References for distances:
$^a$\citet{Mendez2002}, $^b$\citet{Thim2003}, $^c$\citet{Carignan1985}, $^d$\citet{Grocholski2008}, $^e$\citet{Tosi2001}
}
\label{tab:data}
\renewcommand{\footnoterule}{}
\begin{tabular}{lcccccccccccc} \hline\hline
Cluster & \multicolumn{2}{c}{R.A. \& Decl. (2000.0)} & D [Mpc] & $V$ & $A_B$ & log age & Mass [$M_\odot$] & $M_{TO} [M_\odot]$ & FWHM & $\eta$ \\ \hline
\object{NGC1569-A} & 04:30:48.08 & $+$64:50:57.2 & 3.4$^d$ & 14.8 & 2.30 & 6.7 & $6.6\times10^5$ & $\sim25$ & $0\farcs057$/0.9 pc & 1.12 \\
\object{NGC1569-B} & 04:30:48.88 & $+$64:50:51.2 & 3.4$^d$ & 15.6 & 2.30 & 7.2 & $1.2\times10^6$ & $\sim12.0$ & $0\farcs084$/1.4 pc & 1.04 \\
\object{NGC1705-1} & 04:54:13.48 & $-$53:21:39.4 & 5.1$^e$ & 14.8 & 0.04 & 7.1 & $9.2\times10^5$ & $\sim12.5$ & $0\farcs036$/0.9 pc & 1.0 \\
\object{NGC1313-F3-1} & 03:17:47.76 & $-$66:30:18.7 & 4.1$^a$ & 17.1 & 0.47 & 7.74 & $2.8\times10^5$ & 5.9 & $0\farcs19$/3.8 pc & 0.83 \\
\object{NGC5236-F1-1} & 13:37:01.37 & $-$29:50:49.2 & 4.5$^b$ & 17.1 & 0.28 & 7.44 & $2.1\times10^5$ & 7.9 & $0\farcs14$/3.1 pc & 1.09 \\
\object{NGC5236-F1-3} & 13:37:04.67 & $-$29:50:35.2 & 4.5$^b$ & 18.1 & 0.28 & 7.45 & $8.1\times10^4$ & 7.9 & $0\farcs15$/3.3 pc & 1.13 \\
\object{NGC7793-F1-1} & 23:57:35.88 & $-$32:35:40.9 & 3.3$^c$ & 17.4 & 0.07 & 7.69 & $9.2\times10^4$ & 6.3 & $0\farcs23$/3.7 pc & 1.49 \\
\hline
\end{tabular}
\end{minipage}
\end{table*}

Even with the spatial resolution offered by HST, we cannot resolve individual stars in clusters more distant than a few Mpc. For this work, we inspected images taken as part of our Cycle 12 programme (GO-9774, P.I. S. S. Larsen). We found potentially suitable clusters in three out of five galaxies observed in this programme (\object{NGC~1313}, \object{NGC~5236} and \object{NGC~7793}), while no suitable clusters were found in the two other galaxies (\object{NGC~45}, \object{NGC~4395}). The images obtained under this programme were all taken with the wide field channel (WFC) of the ACS, which has a pixel scale of $0\farcs050$ pixel$^{-1}$. At the typical distances of $\sim4$ Mpc, this corresponds to about 1 pc pixel$^{-1}$ so it is clear that we cannot obtain photometry in the cluster cores but only in the outer regions. The filters used for the ACS/WFC observations were F435W, F555W and F814W, roughly equivalent to the Johnson-Cousins $BVI$ filters. For a detailed description of the data, see \citet{Mora2009}.  In addition, we include images of the clusters \object{NGC~1569-A}, \object{NGC~1569-B} and \object{NGC~1705-1}, taken with the ACS/HRC in the filters F330W, F555W and F814W as part of programme GTO-9300 (P.I.: H.\ Ford). The F330W data are less suitable for stellar photometry due to lower S/N (especially for the heavily reddened NGC~1569), and we only make use of the F555W and F814W photometry in this paper. We will from now on refer to the ACS F330W, F435W, F555W and F814W bands as $U$, $B$, $V$ and $I$, although we do not actually transform the photometry to the standard Johnson-Cousins system.

Table~\ref{tab:data} lists basic data for the clusters we have examined in detail. Coordinates were measured on the ACS images; we caution that the absolute HST astrometry is only accurate to $1\arcsec$-$2\arcsec$. Ages for the clusters in \object{NGC~1313}, \object{NGC~5236} and \object{NGC~7793} were estimated from integrated (ground-based) $UBV$ photometry \citep{Larsen1999}, using the empirical calibration of $UBV$ colour vs.\ age in \citet{Girardi1995} which we note is strictly valid only for LMC-like metallicity. The clusters in \object{NGC~1569} and \object{NGC~1705} are significantly younger than those in the spirals, and age estimates from integrated light are more uncertain for these clusters due to the rapid, metallicity-dependent changes in integrated colours at very young ages. For \object{NGC~1705-1} we initially adopt an age of 12 Myr, based on STIS spectroscopy \citep{Vazquez2004}. For \object{NGC~1569-A} we assume an age of 5 Myr \citep{Origlia2001,Westmoquette2007} and for \object{NGC~1569-B} we assume 15 Myr \citep{Anders2004,Larsen2008a}.
The cluster masses were determined as a byproduct of the synthetic cluster experiments described below (Section~\ref{subsec:synt}). We also list the main sequence turn-off mass ($M_{TO}$) corresponding to the estimated age of each cluster, as indicated in the Padua isochrone tables \citep{Bertelli2009}. Since these refer to the bluest point reached during the core H-burning phase, the most massive stars still present in a cluster will be somewhat more massive.

\subsection{Photometry}
\label{sec:photo}

\begin{figure*}
\centering
\includegraphics[width=170mm]{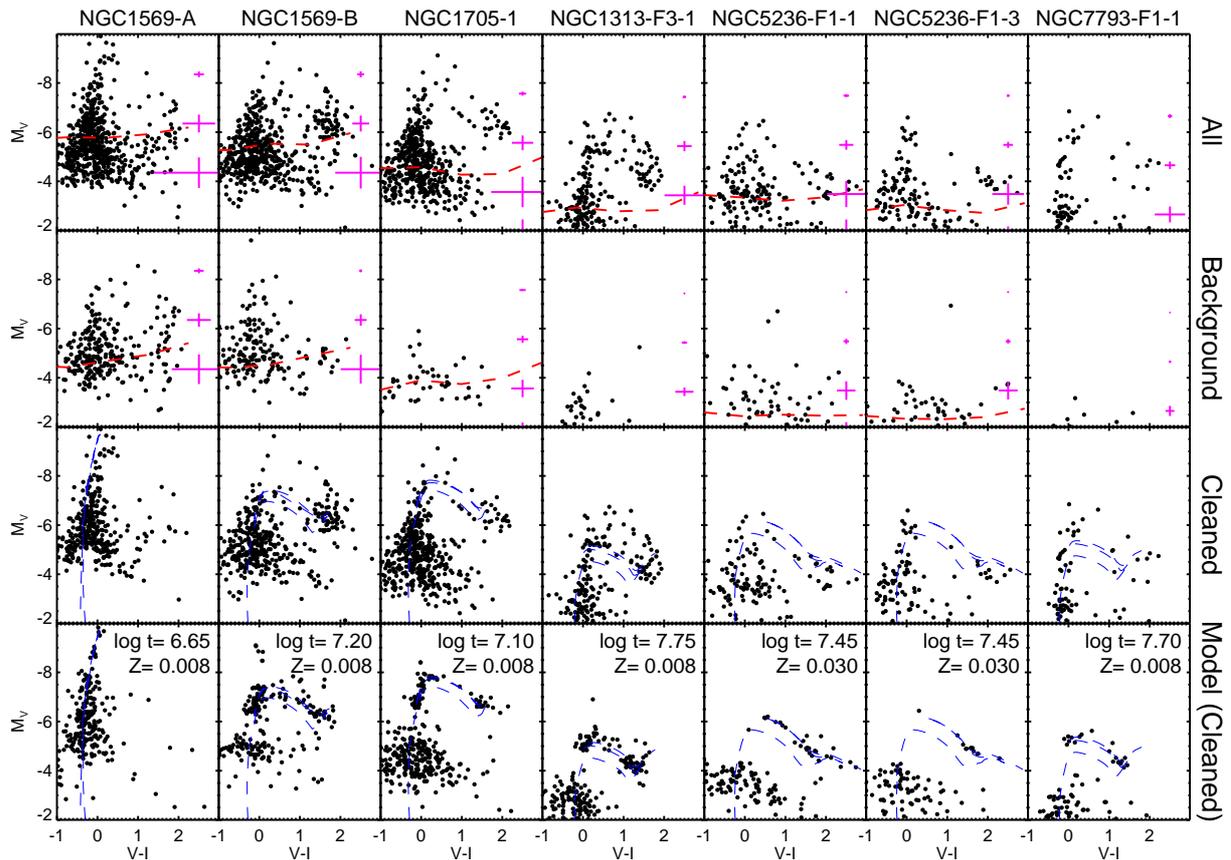}
\caption{\label{fig:cmdvi}Top row: observed $(V\!-\!I), M_V$ colour-magnitude diagrams for the seven clusters, showing all stars for which photometry was obtained. The red dashed curves indicate the approximate 50\% completeness limits, while error bars show typical photometric errors.  Second row from top: Colour-magnitude diagrams for background annulus. Third row:  Background-subtracted CMDs (see text for details). Bottom row: simulated CMDs, using \citet{Bertelli2009} isochrones for the ages and metallicities listed in the legends.}
\end{figure*}

 \begin{figure}
 \centering
 \includegraphics[width=85mm]{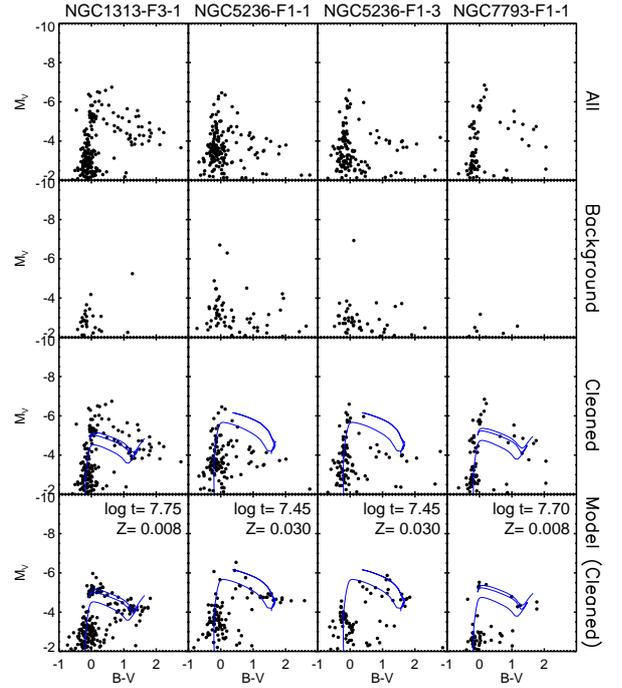}
\caption{\label{fig:cmdbv}Similar to Fig.~\ref{fig:cmdvi}, but showing $(B\!-\!V), M_V$ for those clusters where F435W band data are available.}
\end{figure}

PSF-fitting photometry was done with the DAOPHOT/ALLFRAME package \citep{Stetson1994}. First, the FIND task in DAOPHOT was run on $1000\times1000$ pixels sections of the F555W images, centered on the clusters, and aperture photometry was obtained in all three bands with the PHOT task. Point-spread functions (PSFs) were constructed for each band based on  typically 10--20 stars, and ALLFRAME was then run on all three images simultaneously to produce PSF-fitting magnitudes. In a second iteration of this process, improved PSFs were generated by subtracting all stars except the PSF stars from the images and redetermining the PSFs, and the FIND task was applied once again on the star-subtracted F555W images. Finally, ALLFRAME was run with the improved PSFs and combined  object lists from the first and second iterations as input. The photometry was calibrated by obtaining aperture photometry of the PSF stars on the cleaned images and applying (small) offsets to the PSF magnitudes to make them match the aperture photometry. The photometry was calibrated to the VEGAMAG system using zero-points published on the ACS web pages. We corrected for foreground extinction according to \citet{Schlegel1998}, except for NGC~1569 where we assumed $A_B=2.30$ mag \citep{Israel1988}.

Photometry  was extracted in a circular region with a radius of 2$^{\prime\prime}$ around each cluster, excluding the region within 5 pixels of the centre ($0\farcs25$ for the WFC and $0\farcs125$ for the HRC) where crowding gets too severe to obtain useful photometry. Given that we are generally measuring stars located in the outskirts of the clusters, we expect some contamination of the cluster colour-magnitude diagrams by field stars. In order to correct for this, we used a large comparison annulus with an inner radius of typically $7\farcs5$ and with a similar area as the region where cluster CMDs were obtained. For each star in the comparison annulus, the star in the CMD region which had the most similar $B$, $V$ and $I$  magnitudes was removed (allowing a maximum mismatch of 0.5 mag). For clusters where no $B$ data were available, the matching was done using $V$ and $I$ only. 

The top row in Fig.~\ref{fig:cmdvi} shows the $(V-I, V)$ CMDs for  all stars within the central apertures, while CMDs for the comparison annuli are shown in the second row from the top and the decontaminated cluster CMDs are in the third row. Also shown in the third row are theoretical Padua isochrones for the ages given in Table~\ref{tab:data} and metallicities indicated in each panel (see references in Sect.~\ref{sec:simobs}). Equivalent plots are shown in Fig.~\ref{fig:cmdbv} for the  $(B-V, V)$ photometry.
The cleaned CMDs all show a clear excess above the background and one can readily identify the red supergiant stars at $V-I\approx2$ in several clusters (\object{NGC1569-B}, \object{NGC1705-1}, \object{NGC1313-F3-1}, \object{NGC5236-F1-1}, \object{NGC5236-F1-3}, \object{NGC7793-F1-1}). The CMD for \object{NGC1569-A} also shows a clear excess above the background, although this cluster appears too young to host a significant number of red supergiants. In the remainder of this paper we concentrate on the $(V- I, V)$ photometry, which constitutes a homogeneous dataset for the whole sample.

A first rough indication of the 50\% completeness limits (red dashed lines in Fig.~\ref{fig:cmdvi}) and typical photometric errors was obtained by adding artificial stars to the cluster images, randomly distributed within the central aperture, and redoing the photometry. This was done for $V$ magnitudes between 21 and 27 and $V\!-\!I$ colours between $-1$ and $+3$, in steps of 1 magnitude and for 100 stars at each step. We required a minimum separation of 5 pixels between two artificial stars. The completeness fraction was then determined as the fraction of added stars for which a star was recovered within a separation of 1 pixel. Since these artificial stars are added on top of existing images, there is some probability of recovering a star even where none was added. This number was determined by looking for counterparts to the artificial stars in the original input images and was typically on order $\sim10$\%. An alternative approach would be to simply compare the number of artificial stars added to an image with the additional number of detections. The problem with this approach is that a pre-existing star might be ``masked'' by an artificial star, leading to an underestimate of the number of recovered stars. It is clear that the classical artificial-star method is less than ideal for our situation and below we will adopt a more realistic approach to deal with the severe crowding in our fields.
The classical artificial star approach was also applied to the background annuli (see Fig.~\ref{fig:cmdvi}).  These are less affected by incompleteness than the central apertures. Given the approximative nature of these completeness determinations, we have not attempted to take the different degrees of incompleteness into account when correcting for the background contribution, so we are likely over-subtracting it. However, from Fig.~\ref{fig:cmdvi} and \ref{fig:cmdbv} it is clear that this mainly affects the CMDs near the completeness limits where errors are large anyway. The post-main sequence stars are, in most cases, hardly affected by the background correction, and well above the 50\% completeness limits.

\subsection{Synthetic cluster experiments}
\label{subsec:synt}

\begin{figure}
\centering
\includegraphics[width=85mm]{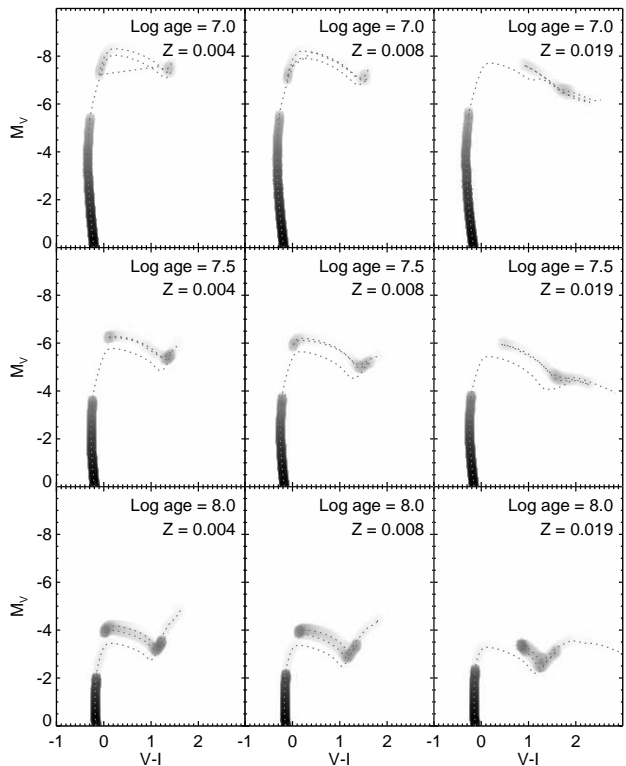}
\caption{\label{fig:isocmp_p08}Hess diagrams for Padua isochrones \citep{Bertelli2009} of various ages and metallicites. The dotted line in each panel connects each point on the isochrones, while the shaded areas indicate the density of stars (assuming a Kroupa IMF).}
\end{figure}

\begin{figure}
\centering
\includegraphics[width=85mm]{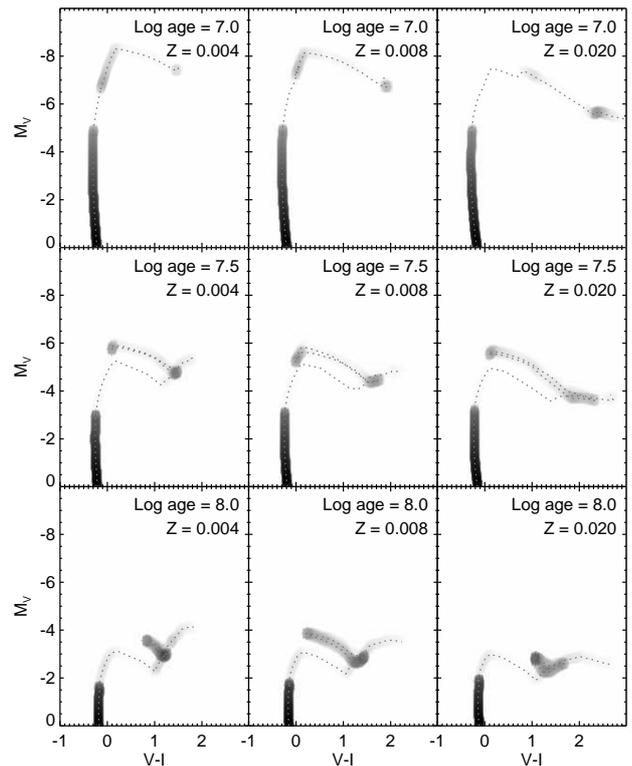}
\caption{\label{fig:isocmp_g01}Same as Fig.~\ref{fig:isocmp_p08}, but for Geneva isochrones \citep{Lejeune2001}. Note that the Geneva isochrones are shown for WFPC2 filter definitions.}
\end{figure}

A simple comparison with isochrones is of limited use for interpreting data such as those in Fig.~\ref{fig:cmdvi}, since this fails to take into account the strongly varying density of stars along a particular isochrone. More information can be gained by plotting the isochrones as Hess diagrams, as shown in Fig.~\ref{fig:isocmp_p08} (for the Padua isochrones) and Fig.~\ref{fig:isocmp_g01} \cite[for the Geneva isochrones; ][]{Lejeune2001}. It should be noted that the Geneva isochrones are shown for the HST \emph{Wide Field / Planetary Camera 2} F555W and F814W filters, which are not completely identical to the filters with the same names on the ACS. However, for the purpose of illustration the differences are marginal.  

The main characteristics of the Padua and Geneva isochrones are similar, although there are differences in the details. The predicted Blue Hertzsprung Gap is evident at all ages relevant to the data analysed here. It is already clear from Fig.~\ref{fig:cmdvi} that this feature is not nearly as evident in the data. In what follows, we will restrict the comparison to the Padua isochrones, for which photometry in the ACS filters is directly available.

\begin{figure*}
%
\begin{minipage}{180mm}
\makebox[25mm]{NGC1569-A}
\makebox[25mm]{NGC1569-B}
\makebox[25mm]{NGC1705-1}
\makebox[25mm]{NGC1313-F3-1}
\makebox[25mm]{NGC5236-F1-1}
\makebox[25mm]{NGC5236-F1-3}
\makebox[25mm]{NGC7793-F1-1}
\end{minipage} \\
\begin{minipage}{180mm}
\includegraphics[width=25mm]{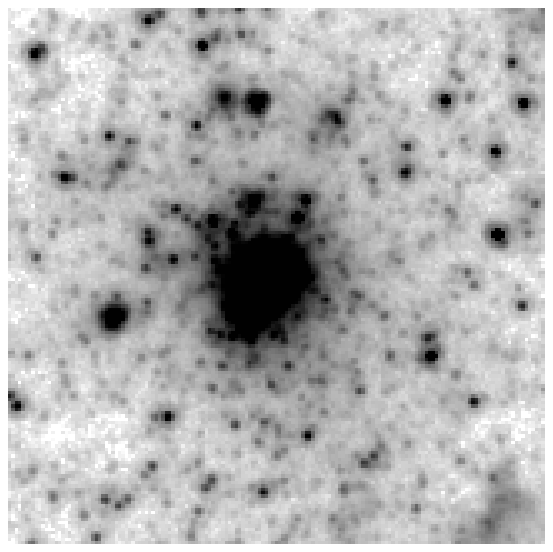}
\includegraphics[width=25mm]{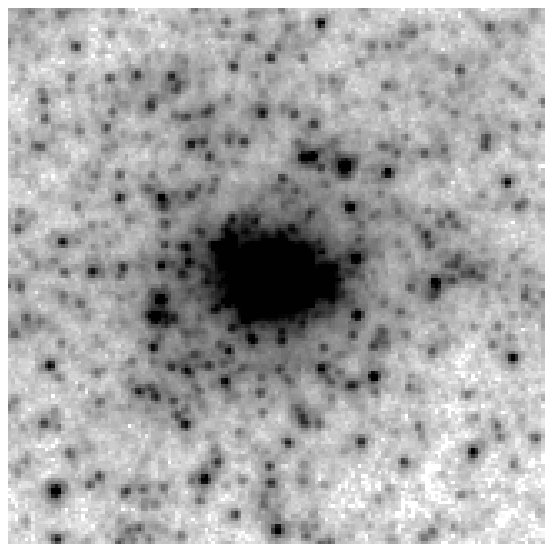}
\includegraphics[width=25mm]{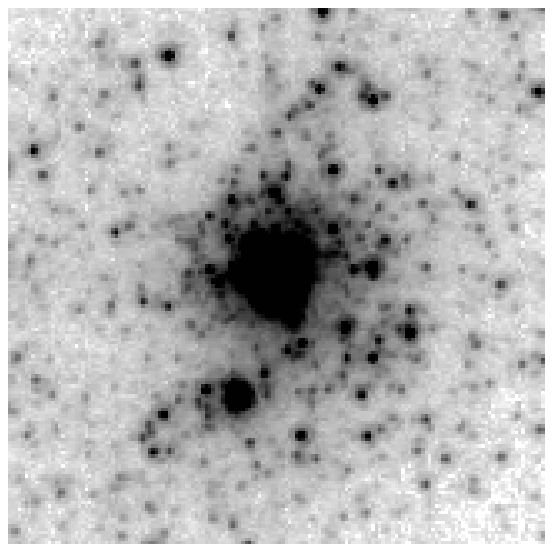}
\includegraphics[width=25mm]{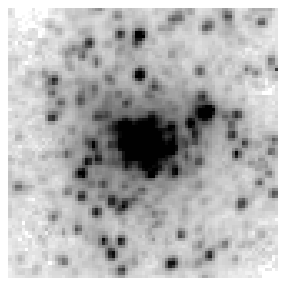}
\includegraphics[width=25mm]{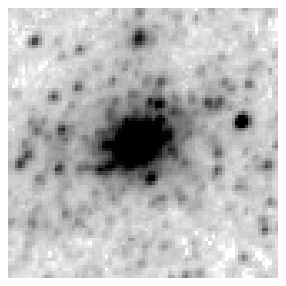}
\includegraphics[width=25mm]{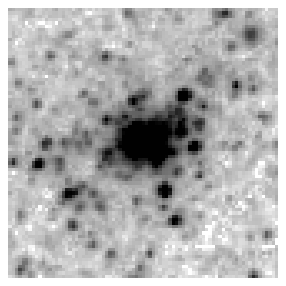}
\includegraphics[width=25mm]{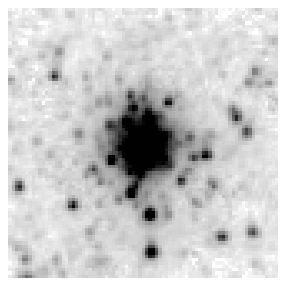}
\end{minipage} \\
\begin{minipage}{180mm}
\includegraphics[width=25mm]{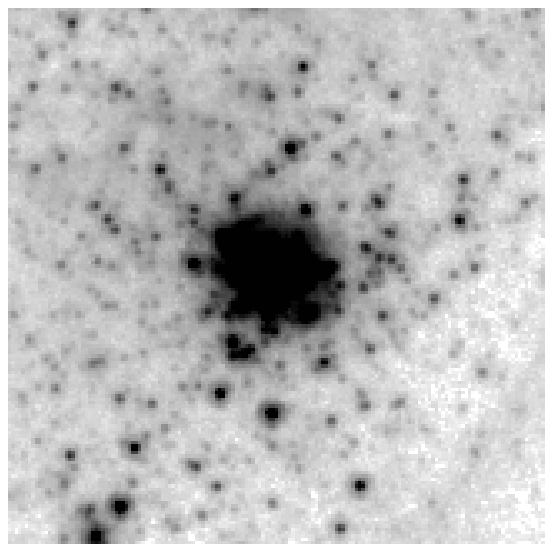}
\includegraphics[width=25mm]{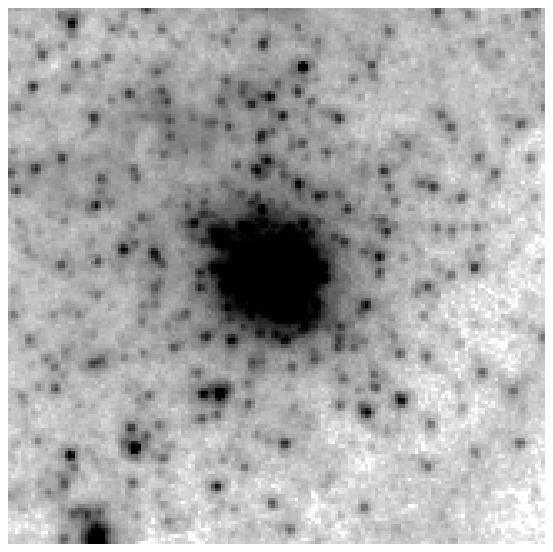}
\includegraphics[width=25mm]{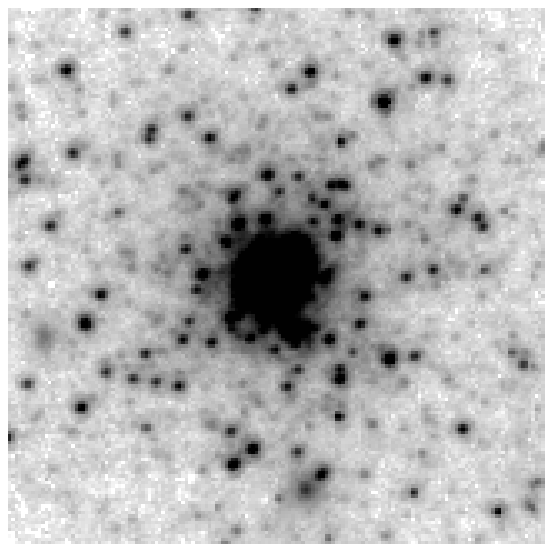}
\includegraphics[width=25mm]{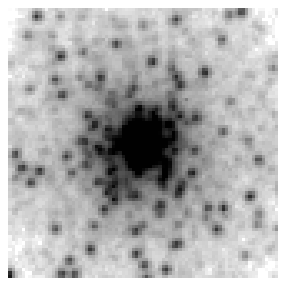}
\includegraphics[width=25mm]{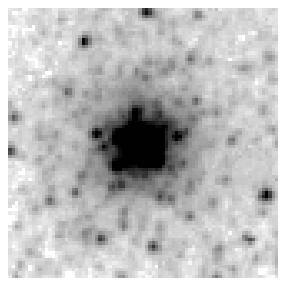}
\includegraphics[width=25mm]{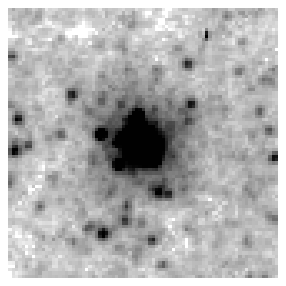}
\includegraphics[width=25mm]{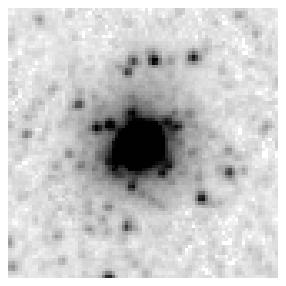}
\end{minipage}
\caption{\label{fig:sim_real}Top panels: F555W images of the star clusters. Bottom panels: simulated clusters with the same apparent magnitudes, ages and sizes. Each panel measures $4^{\prime\prime} \times 4^{\prime\prime}$.}
\end{figure*}

Even though the outskirts of the clusters appear resolved into individual stars in the ACS images, crowding still represents a major challenge and we cannot directly compare our observed CMDs with the Hess diagrams in Fig.~\ref{fig:isocmp_p08} and \ref{fig:isocmp_g01}. A realistic assessment of photometric errors and detection limits is crucial in order to carry out a meaningful comparison between the observed CMDs and model predictions. As we have discussed above, the classical approach of adding stars at random positions within an image is insufficient to fully capture the complicated effects of crowding in our data. In addition to the effects mentioned in Sec.~\ref{sec:photo}, the background and crowding vary strongly with distance from the cluster centre, and a different approach is needed. Rather than relying on ``typical'' completeness limits and photometric errors, we therefore opted to generate artificial counterparts of the real cluster images, and carry out photometry on those. In this way, we could directly compare the CMDs of our \emph{artificial} clusters with those of the actual clusters.

To generate the artificial clusters, we first derived structural parameters for each real cluster using the ISHAPE task in the BAOLAB package \citep{Larsen1999}. ISHAPE assumes an analytic model of each cluster, 
where for this work we use profiles of the following type, found by \citet{Elson1987} to give a good fit to clusters in the Large Magellanic Cloud:
\begin{equation}
  I(r) = I_0 \left[1 + \left(\frac{r}{r_c}\right)^2\right]^{-\eta}.
  \label{eq:eff}
\end{equation}
Here, $r_c$ is a core radius and $\eta$ determines how steeply the envelope falls off at large radii. ISHAPE convolves the analytic cluster model with the PSF and iteratively adjusts the structural parameters of the cluster model until the best match to the data is achieved. In our case, the signal-to-noise was sufficiently high to allow ISHAPE to solve for both $r_c$ and $\eta$. The last two columns in Table~\ref{tab:data} list the best-fitting $\eta$ and FWHM (Full Width at Half Maximum) for each cluster, the latter related to $r_c$ and $\eta$ as
\begin{equation}
  {\rm FWHM} = 2 \, r_c \, \sqrt{2^{1/\eta - 1}}
\end{equation}
for the \citet{Elson1987} profiles. For the clusters located in spiral galaxies (\object{NGC 1313}, \object{NGC 5236}, \object{NGC 7793}) the FWHM values are fairly typical compared to other clusters of similar masses in spirals \citep{Larsen2004}, while the clusters in \object{NGC~1569} and \object{NGC~1705} have extraordinarily compact cores. 
This is particularly remarkable given that the latter are also the most \emph{massive} clusters in our sample, implying very high stellar densities in the centre of these clusters ($\ga 10^5 M_\odot$ pc$^{-3}$).
We do not tabulate half-light radii, which are very uncertain because the $\eta$ values are close to unity for most of the clusters (the half-light radii are, in fact, undefined for $\eta\leq1$, unless the profiles are truncated at some finite outer radius).


Having determined the intrinsic intensity profile of each cluster, we then proceeded to sample stars at random from the model profiles. The profiles in Eq.~(\ref{eq:eff}) extend to infinity, but we imposed a cut-off at a radius of 50 pixels ($2\farcs5$ for the WFC images and $1\farcs25$ for the HRC images). 
Each star was assigned a mass drawn at random from a \citet{Kroupa2002} IMF, and corresponding magnitudes in each band were looked up in the Padua isochrones. Isochrones were selected according to the cluster ages given in Table~\ref{tab:data} and the total number of stars in each artificial cluster was tuned such that the integrated $V$-band magnitudes matched those of the real clusters.
The stars belonging to these artificial clusters were then added, one by one, to the actual images using the MKSYNTH task in BAOLAB. To the extent possible, we added the artificial clusters at locations that resembled those of the real clusters in terms of the background level and stellar density. 
The photometric analysis described in Sec.~\ref{sec:photo} was then repeated on the images with the artificial clusters added. 

Real and simulated F555W images of each cluster are shown in Fig.~\ref{fig:sim_real}. 
Although our simulated clusters closely resemble the real clusters, a couple of caveats should be noted: First, while we have taken great care to model the photometric (in)completeness and errors as realistically as possible using simulated images, there are always simplifications involved in the generation of artificial datasets. For example, the PSFs used in the artificial cluster images are approximations to the actual PSFs of the parent images, so that stars in the artificial clusters will have slightly different PSFs than the real stars present in the images. This may introduce subtle differences in the goodness of the ALLFRAME fits, possibly affecting the photometric errors. This would be a concern especially if we used the same PSFs to generate artificial clusters \emph{and} carry out PSF-fitting photometry for them.  To alleviate this concern and mimic the process by which PSFs were derived for real clusters as closely as possible, we re-derived the PSFs for the synthetic images using a set of synthetic reference stars. These were placed in similar locations within the images as the real PSF stars in terms of crowding and background level. Another potential concern is our implicit neglect of mass segregation. This effect causes massive stars to move towards the centre of a cluster and low-mass stars to move to the outskirts. However, since the light profiles of both the artificial and real clusters are dominated by the massive (luminous) stars, we do not expect this simplification to affect our analysis very strongly. Potentially more problematic is our neglect of binary stars. While random blends are automatically accounted for, physical binaries are not. This is a complex problem: a full account of binaries needs to consider not just the blending of light from two stars, but also possible effects of stellar evolution in binary systems (mass transfer, etc.) for realistic distributions of mass ratios and orbital periods. We will comment further on some possible effects of binary evolution on the CMDs below (Sect.~\ref{sec:binaries}).

\subsection{Simulated versus observed CMDs}
\label{sec:simobs}

\begin{figure*}
\begin{minipage}{180mm}
\begin{minipage}{44mm}
\includegraphics[width=44mm]{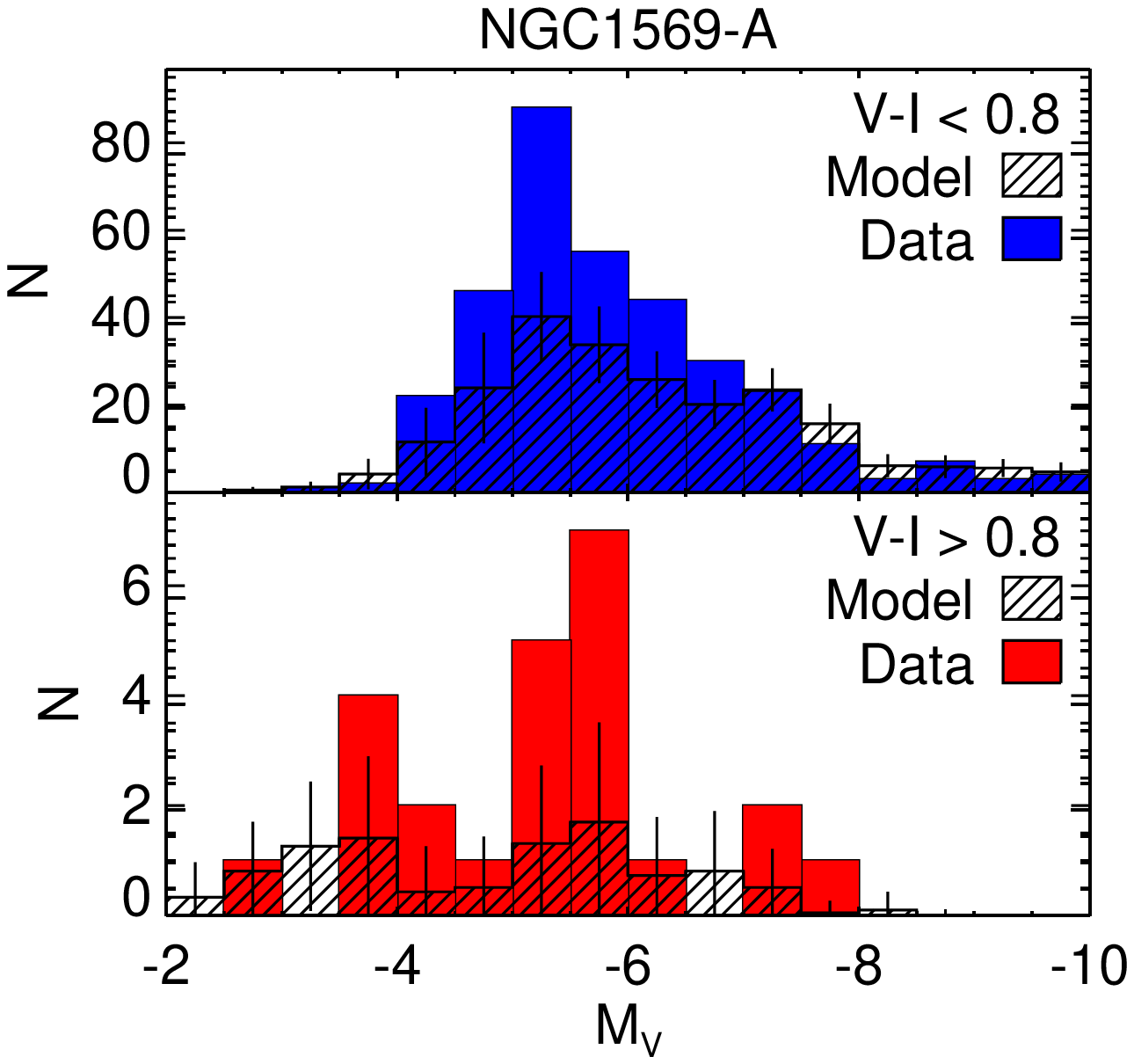}
\end{minipage}
\begin{minipage}{44mm}
\includegraphics[width=44mm]{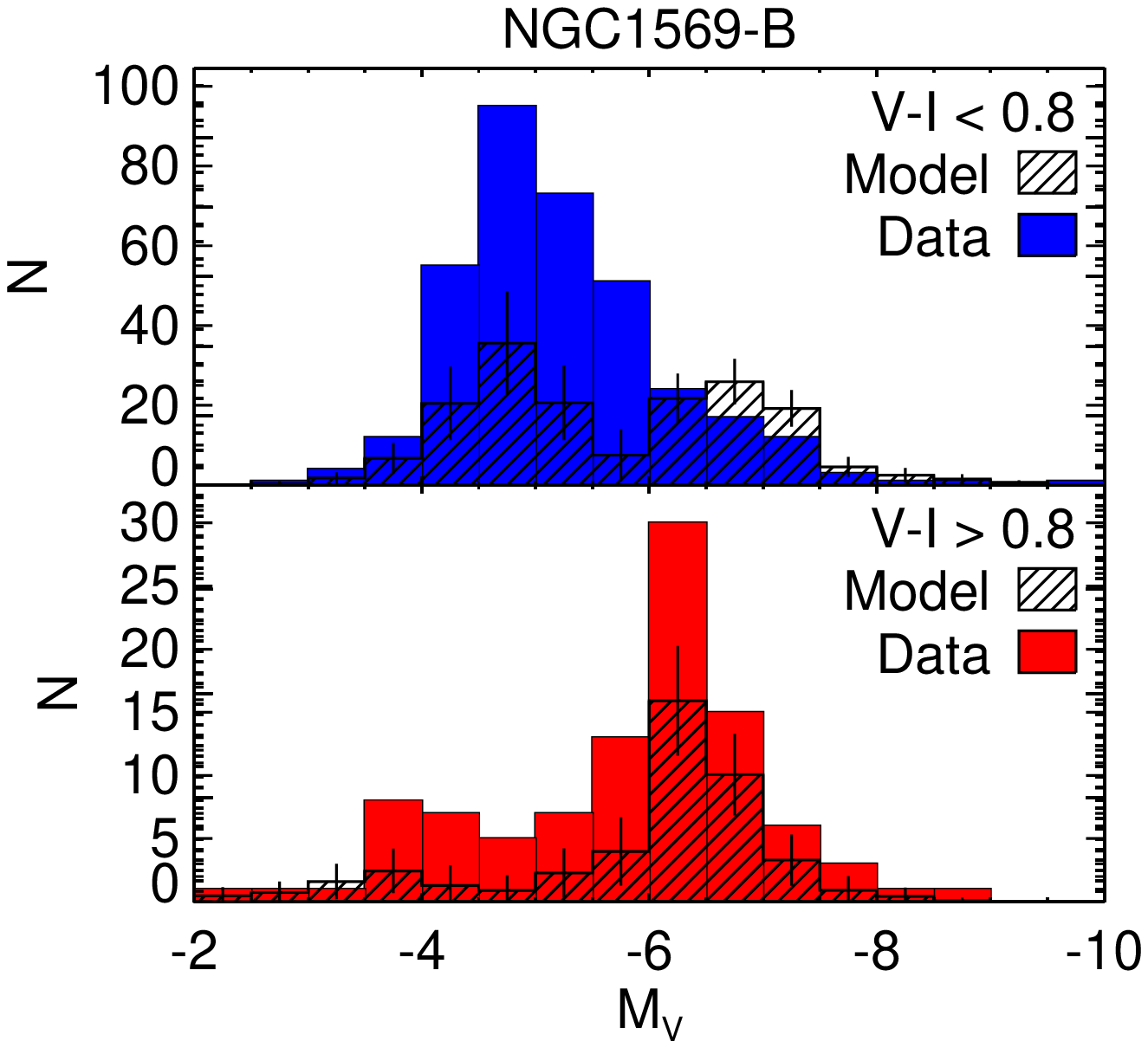}
\end{minipage}
\begin{minipage}{44mm}
\includegraphics[width=44mm]{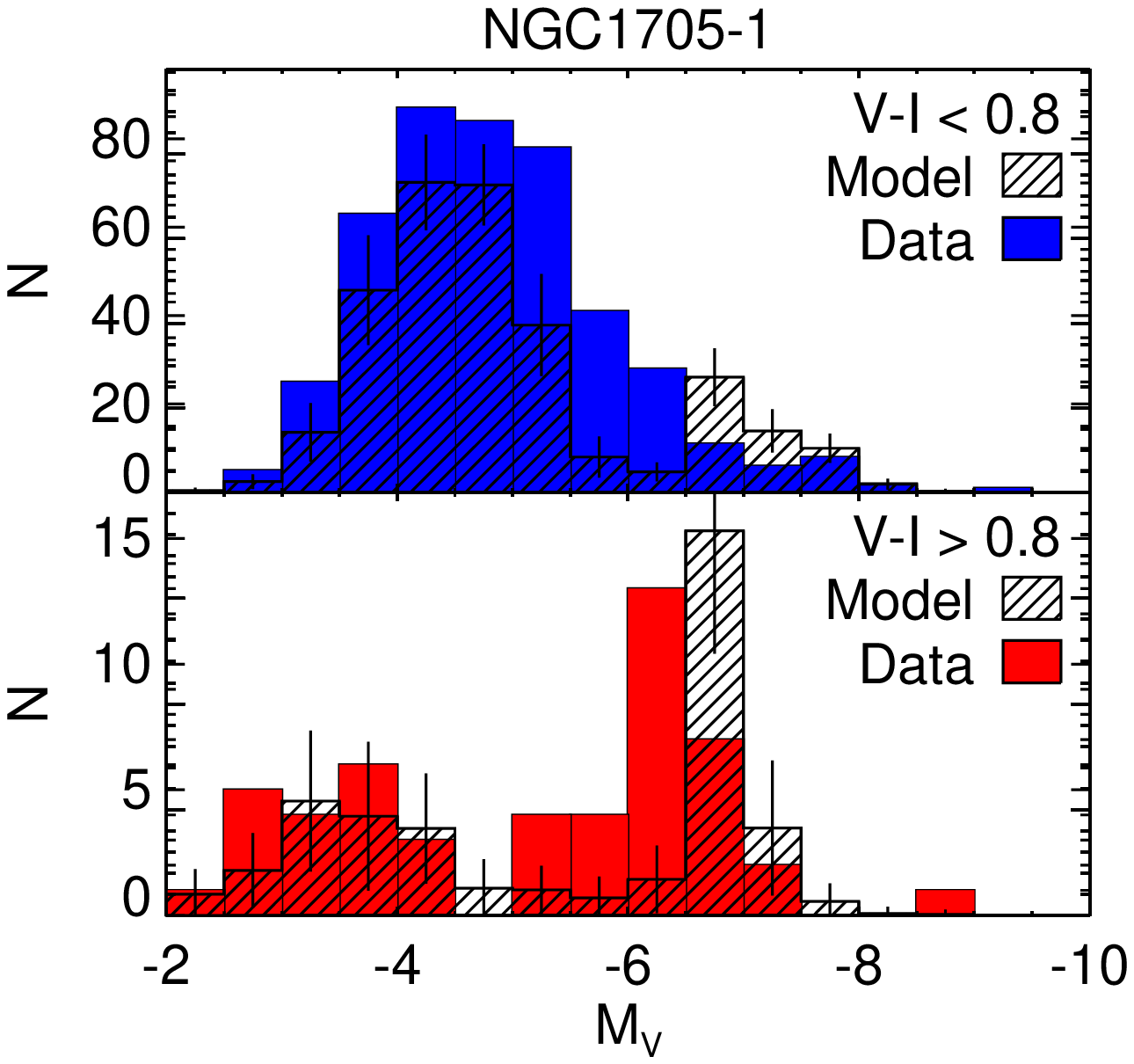}
\end{minipage}
\begin{minipage}{44mm}
\includegraphics[width=44mm]{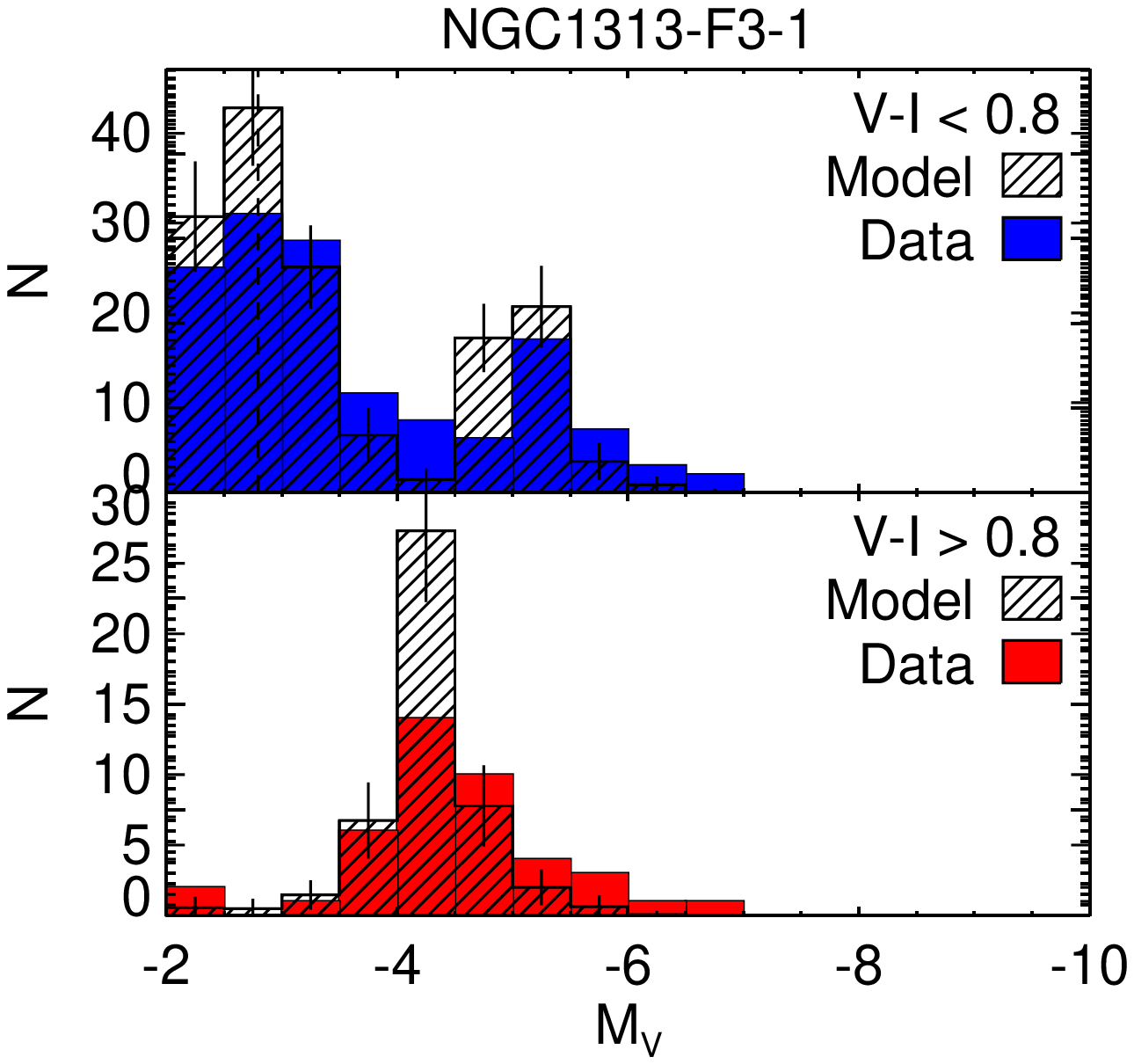}
\end{minipage}
\end{minipage}
\begin{minipage}{180mm}
\begin{minipage}{44mm}
\includegraphics[width=44mm]{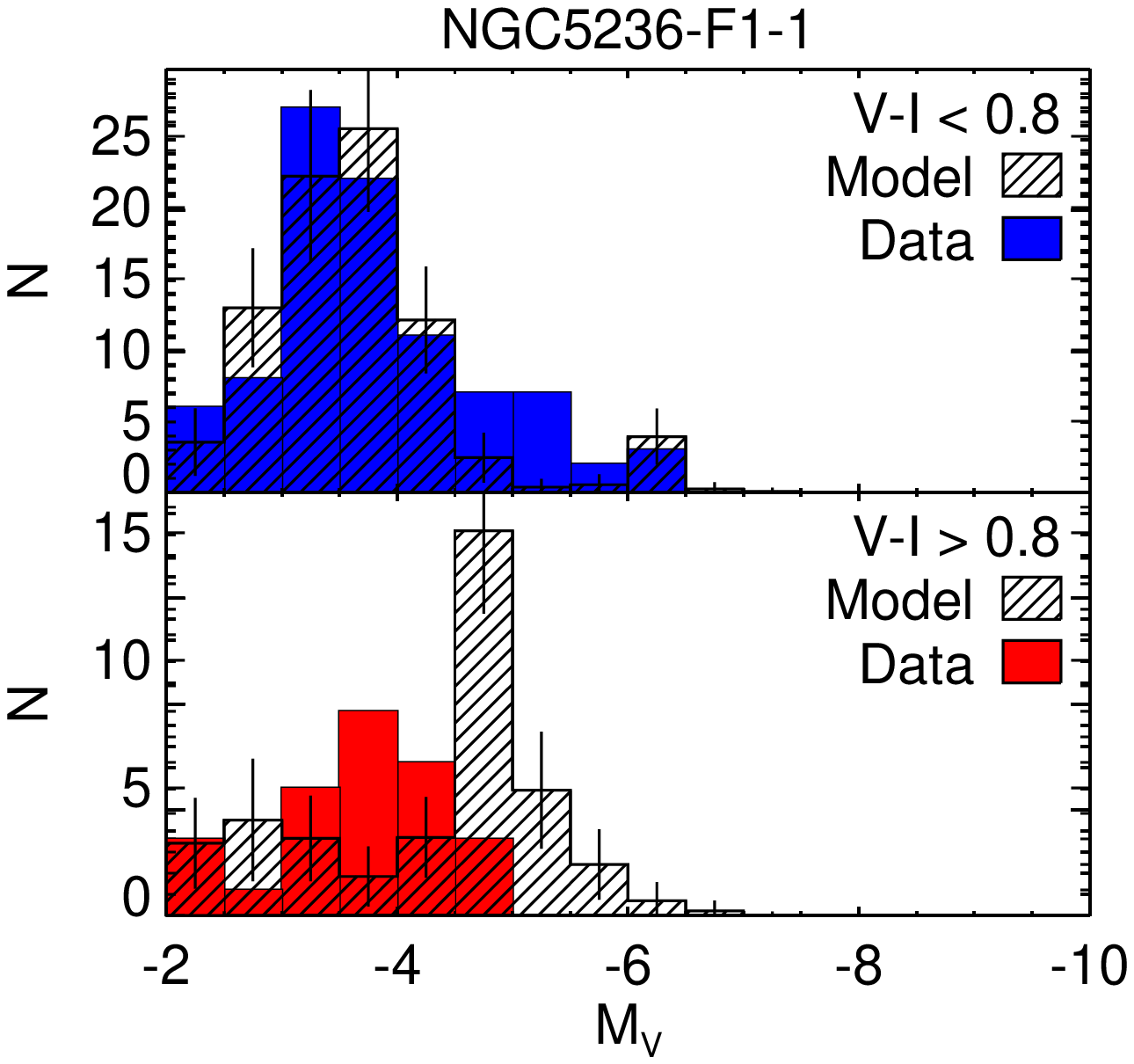}
\end{minipage}
\begin{minipage}{44mm}
\includegraphics[width=44mm]{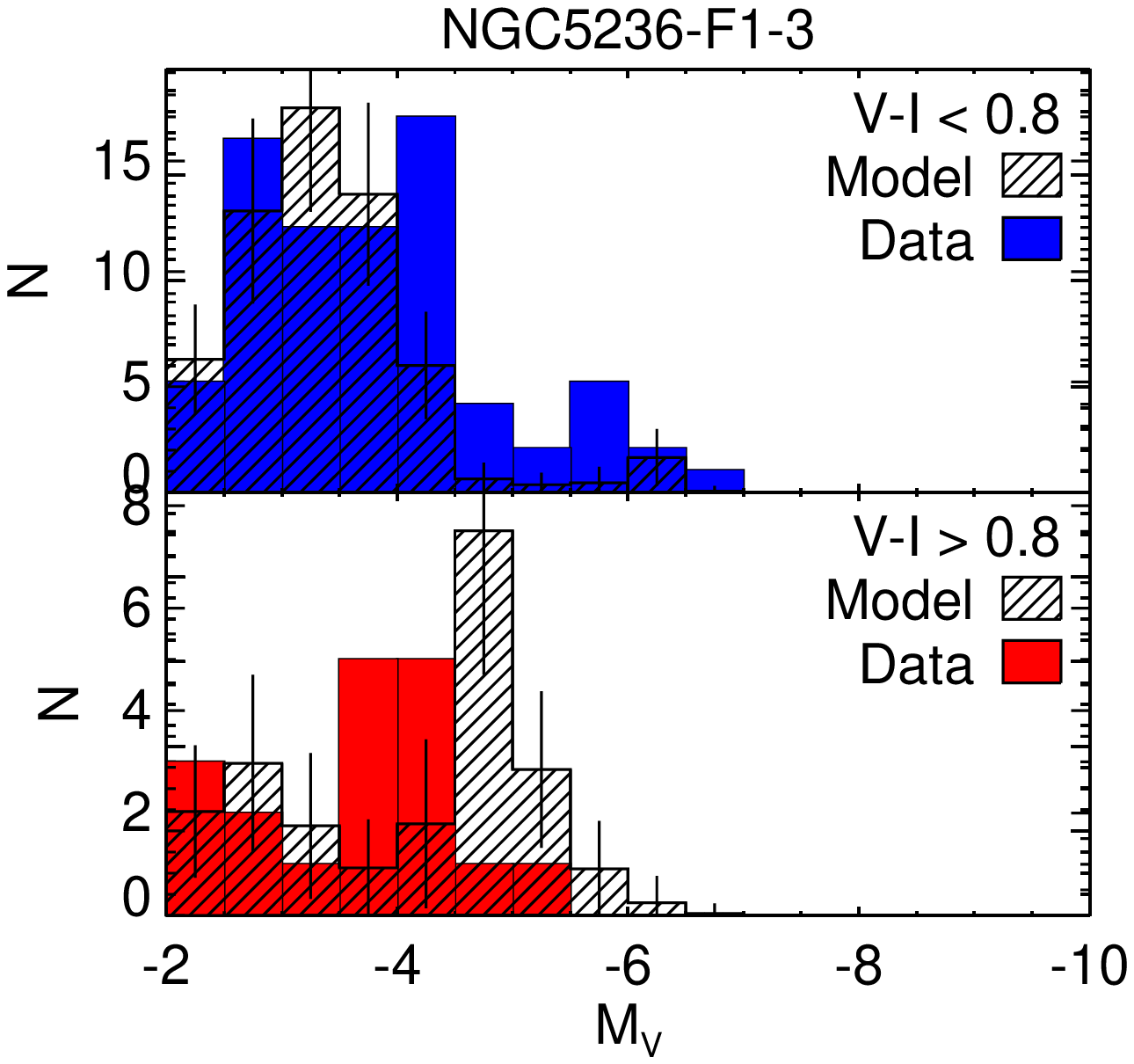}
\end{minipage}
\begin{minipage}{44mm}
\includegraphics[width=44mm]{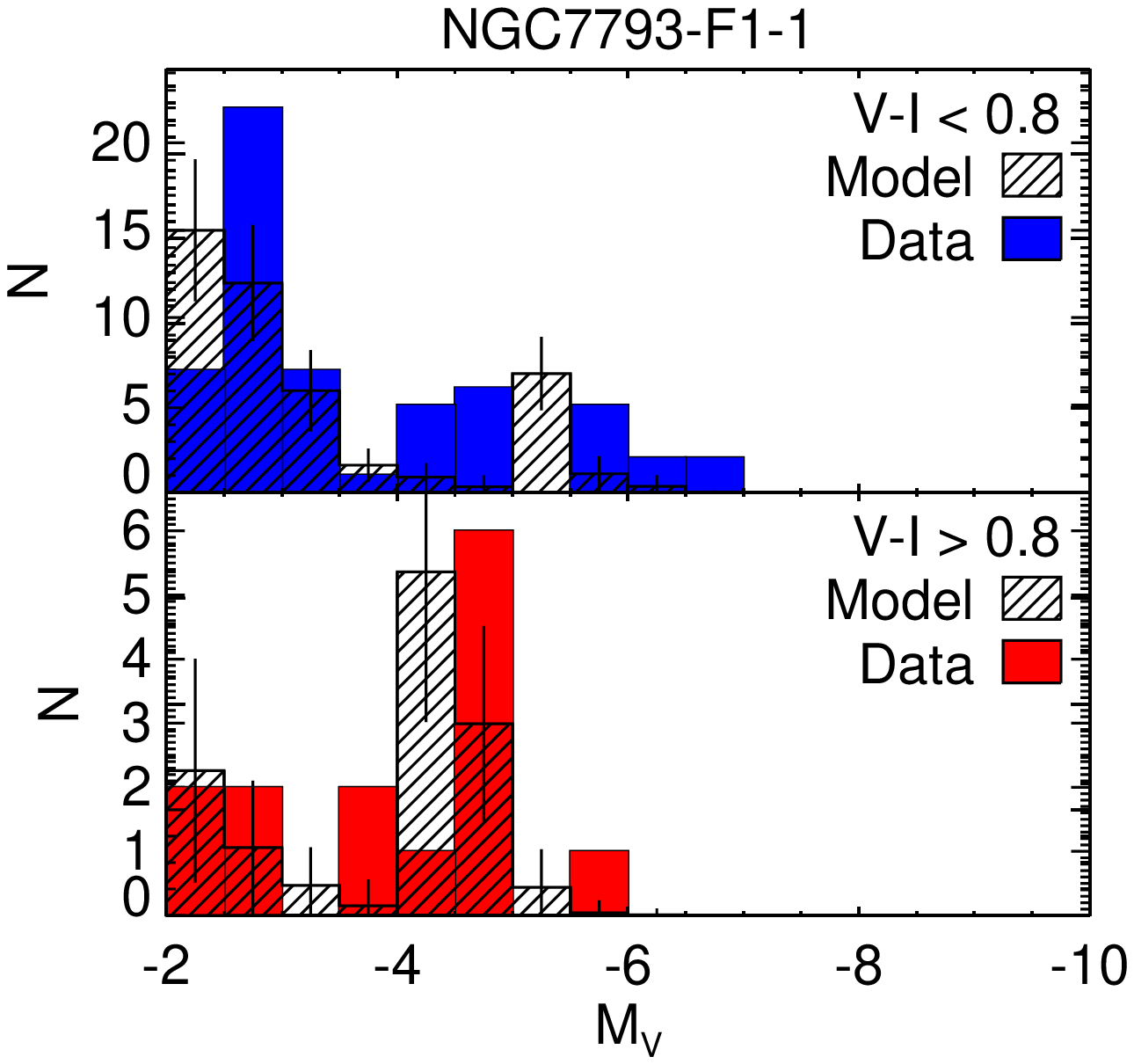}
\end{minipage}
\end{minipage}

\caption{\label{fig:lfs}Observed and simulated luminosity functions for red and blue stars in the clusters.}
\end{figure*}

CMDs for the simulated clusters are shown in the bottom rows of Fig.~\ref{fig:cmdvi} and Fig.~\ref{fig:cmdbv}. Background contamination has been statistically subtracted in the same way as for the real clusters.
Also shown are the isochrones used to generate the lists of input magnitudes. 
We have somewhat arbitrarily, but guided by the CMDs in Fig.~\ref{fig:cmdvi}, made a division between ``red'' and ``blue'' stars at $V-I = 0.8$
In Fig.~\ref{fig:lfs} we compare the observed and model luminosity functions for blue and red  stars. The ``turn-overs'' in both the simulated and observed luminosity functions generally agree fairly well with the approximate 50\% completeness limits indicated in Fig.~\ref{fig:cmdvi}.

For the artificial clusters, both the LFs and the synthetic CMDs are averages of 100 individual realizations of each cluster. For each realization, a different random number seed was used for the sampling of the IMF and the cluster profile, and the centre of the cluster was shifted around randomly within a box of $\pm50$ pixels. This approach allows us to put one-sigma error bars on the simulated LFs that include not only stochastic fluctuations in the clusters themselves, but also background variations. It should once again be emphasized that the luminosity functions have not been corrected for completeness effects, so we do not attempt to derive quantitative estimates of the shapes of these distributions. Instead, we keep the comparison with the simulated data strictly differential.
Similarly, the simulated CMDs shown in the figures are ``average'' CMDs generated by first combining the 100 individual CMDs for each artificial cluster, and then randomly sampling 1\% of the stars.

Even though we employ PSF-fitting photometry, there is still the possibility that chance superpositions of two or several stars (as opposed to physical binaries) overlap so closely that they will be fitted as a single star. This is likely responsible for some of the data points in the synthetic CMDs that fall significantly off the input isochrones. The effect will be present both in the simulated and real data, and should not affect the relative comparison. Using our synthetic clusters, we can obtain some rough estimates of the amount of blending. The percentage of stars affected by blending will clearly depend on the specific criterion for a blend to occur. If we (somewhat arbitrarily) define a ``blend'' as a star with a companion that is closer than 1 pixel and at most one magnitude fainter than the primary star, we find that roughly 20\% of the brightest stars in our CMDs are affected by blending. However, this fraction depends very strongly on position and magnitude.

In terms of the magnitude limits and scatter in the colour of main sequence stars, the simulated CMDs resemble the observed CMDs quite well, giving us some confidence that the modelling procedure captures the most important observational effects. However, a detailed comparison shows several significant differences between the modelled and observed CMDs. We will now comment further on these. We first discuss similarities and differences between the observed and model CMDs for each cluster, and then consider how various modifications to our initial model assumptions might affect the model CMDs.

\paragraph{\object{NGC~1569-A}:}
Comparison with the other CMDs clearly shows this cluster to be the youngest in our sample, in agreement with analyses based on integrated light.
It is actually a binary cluster, as already hinted at in pre-refurbishment mission HST \emph{Wide Field / Planetary Camera} data \citep{OConnell1994}. 
In our model CMDs we have assumed a metallicity of $Z=0.008$ for this cluster and \object{NGC~1569-B} \citep[see discussion in][]{Larsen2008a}.
Spectroscopic studies have revealed the presence of both Wolf-Rayet and red supergiant stars in \object{NGC~1569-A} \citep{Origlia2001}.  However, while the raw CMD in the top panel of Fig.~\ref{fig:cmdvi} does show a number of RSGs, the background annulus shows a similar population so we find no statistically significant population of RSGs in \object{NGC~1569-A} itself. 
We do however note that a number of the brightest stars (at $M_V\sim-9$) appear somewhat redder  than any stars appearing in the model CMD. These stars cannot be accounted for by any choice of isochrone: the isochrone used for the modelling (log(age)=6.65) is the oldest one for which no RSGs are present. For log(age)=6.70, the RSGs appear suddenly at $V\!-\!I \sim 1.0$. The observed and modelled LFs do not match particularly well either; on the blue side the simulated LF is significantly shallower than the observed one. The LF for red stars is most likely due to residual background contamination.


\paragraph{\object{NGC~1569-B}:}
The observed CMD reveals a large number of RSGs (about 60) which have been discussed in detail in \citet{Larsen2008a}.  Fig.~\ref{fig:lfs} shows that the shape of the luminosity function of the RSGs is fairly well matched by the simulation for the assumed age and metallicity, although the detailed morphologies of the simulated and observed CMDs do differ somewhat. In particular, the observed luminosities of the RSGs show more scatter at fixed colour than the simulation.
Overall, the simulated CMD appears sparser than the observed one. Although, by construction, the \emph{total} luminosities of the simulated clusters match those of the observed ones, the integrated luminosity of the \emph{measured} stars is about 30\% lower in the simulation for \object{NGC~1569-B}. Thus, the DAOPHOT photometry appears to be slightly less complete for the simulated cluster. This may be due to slight inaccuracies in the modelling of the cluster profiles or differences in the CMDs.

Similarly to several other clusters, the simulated CMD shows a clear BHG which is not present in the real data. This difference is also clear when comparing the observed and simulated LFs. A small number of very luminous stars ($M_V < -8$) are observed; however, these are also present in the simulation indicating that they may well be blends or other artifacts due to the extreme crowding in this very dense cluster.  


\paragraph{\object{NGC~1705-1}:}
The image of this cluster shows considerable substructure, with two ``filaments'' and another fainter cluster located about $1^{\prime\prime}$ from the centre of the main core. The metallicity of \object{NGC~1705} is intermediate between those of the SMC and LMC \citep{Lee2004} and for the simulation we have assumed $Z=0.008$.  The observed CMD of this cluster is very similar to that of \object{NGC~1569-B}, indicating similar ages for the two clusters, and also for \object{NGC~1705-1} the simulated CMD is a rather poor match to the observed one. Again, a BHG is clearly expected at $M_V\sim-6$, but not observed. Fig.~\ref{fig:lfs} shows that the observed and simulated numbers of red supergiants match quite nicely, although the observations show the RSGs to be somewhat cooler and fainter than predicted by the models for this age and metallicity. 


\paragraph{\object{NGC~1313-F3-1}:}
This is a remarkable object. With a core radius of about 4 pc, this is a relatively extended cluster and the HST images show it to be resolved into numerous individual stars. The metallicity of \object{NGC~1313} is 12$+$log O/H = $8.4\pm0.1$ \citep{Walsh1997}, which translates to [O/H] = $-0.43\pm0.1$ \citep{Grevesse1998} or $Z=0.007$ assuming Solar abundance ratios. We thus use the $Z=0.008$ Padua models for the simulated CMD, which to first order results in quite a good match to the data. This is one of the few cases where the observed CMD and LF show a hint of the BHG that is expected at $M_V\sim-4$. However, the observed CMD shows a spread of nearly 2 mag in the $M_V$ magnitudes of the supergiant stars, a significantly larger spread than in the simulated CMDs. 


\paragraph{\object{NGC~5236-F1-1} and \object{NGC~5236-F1-3}:}
The metallicity of most of the \object{NGC~5236} disc is well above Solar \citep{Bresolin2002} and we have used the $Z=0.03$ isochrones for the simulations. However, the simulated CMDs would not look much different for the Solar metallicity isochrones ($Z=0.019$). The integrated colours suggest identical ages for these two clusters, and their observed CMDs are indeed quite similar and in neither case well reproduced by the simulations.  \object{NGC~5236-F1-1} is the brighter of the two by a magnitude, and the CMD shows about twice as many RSGs compared to \object{NGC~5236-F1-3}. The observed $M_V$ magnitudes of the RSGs are about a magnitude fainter than the simulated ones, possibly an indication that the clusters are older than we have assumed. 
The mismatch between models and data is even worse on the blue side of the CMD. The models predict a clear BHG both in $M_V$ magnitude and colour, whereas the data show no such separation. This problem cannot be solved by adopting an older age.


\paragraph{\object{NGC~7793-F1-1}:}
The observed CMD is similar to that for \object{NGC~1313-F3-1} but more sparsely populated, consistent with the $\sim3\times$ lower estimated total mass. We have assumed $Z=0.008$ \citep{Edmunds1984}. Many of the same remarks given for \object{NGC~1313-F3-1} apply to this cluster, but at lower statistical significance. Also here, there appears to be a significant 1--2 mag spread in the $M_V$ magnitudes of the supergiant stars. 

\section{Discussion}

\subsection{General comments on CMD comparisons}

Comparing our observed and simulated CMDs, we first note that they do agree in some respects. The colour of the main sequence is, in general, quite well reproduced, an indication that our adopted extinction values are reasonable. In addition, the CMDs show the red supergiants to be fairly well separated from the blue loop and main sequence stars, in agreement with model predictions for the relevant mass/age range (Fig.~\ref{fig:isocmp_p08} and \ref{fig:isocmp_g01}). However, there are also several areas of disagreement: 1) Even after accounting for observational errors, we confirm the presence of stars in the BHG where none are expected. This is a long-standing problem \citep{Chiosi1998,Salasnich1999} and has previously been noted e.g. for the $\sim50$ Myr old cluster \object{NGC~1850} in the LMC \citep{Fischer1993}. Likewise, inspection of the CMD for the 25-Myr old cluster \object{vdB0} in \object{M31} \citep{Perina2009} shows no clear gap, although statistics are poorer here. 2) In several cases, the observations show a larger spread in the luminosities of the supergiant stars than expected from our artificial-cluster experiments. 3) The observed colours (i.e., effective temperatures) of the red supergiant stars generally tend to be somewhat redder than the predictions. This is seen both at super-solar metallicities (i.e., in NGC~5236), as well as in the more metal-poor (but younger) clusters \object{NGC~1569-B} and \object{NGC~1705-1} \citep[see also][]{Larsen2008a}. The agreement between models and data is best for the clusters \object{NGC~1313-F3-1} and \object{NGC~7793-F1-1}, which are the oldest in our sample and probably have slightly sub-solar metallicities.

In the following we discuss several modifications to our initial, simple model assumptions that might affect the CMDs.

\subsection{Age spreads?}
\label{sec:agespread}

\begin{figure}
\centering
\includegraphics[width=85mm]{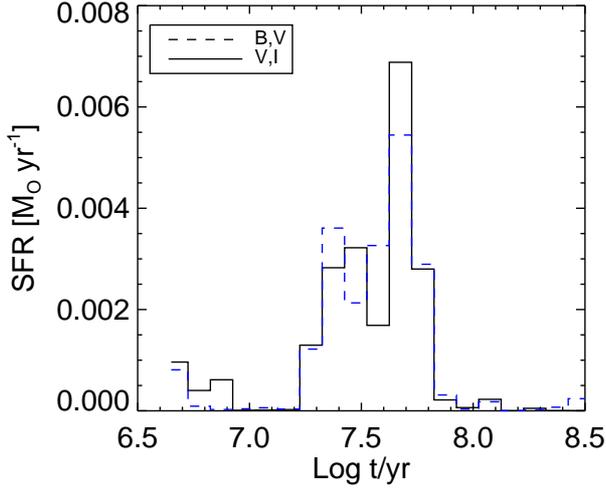}
\caption{\label{fig:sfh1313_f3_1}Star formation history for NGC1313-F3-1 derived via CMD synthesis. Solid curve: fit from $V,I$ colours. Dashed curve: fit from $B,V$.  
 }
\end{figure}

\begin{table}
\begin{minipage}[t]{\columnwidth}
\caption{Input for multiple-burst models. $\log t_1$ - $\log t_3$ indicate the logarithmic ages (in years) of the three bursts and $f_1$ - $f_3$ the relative amounts of stars formed in each burst.
}
\label{tab:mbursts}
\renewcommand{\footnoterule}{}
\begin{tabular}{lcccccc} \hline\hline
Cluster & $\log t_1$ & $\log t_2$ & $\log t_3$ & $f_1$ & $f_2$ & $f_3$ \\ \hline
NGC1569-B & 7.20 & 6.95 & 6.60 & 0.70 & 0.15 & 0.15 \\
NGC1705-1 & 7.25 & 7.05 & 6.60 & 0.45 & 0.25 & 0.30 \\
NGC1313-F3-1 & 7.70 & 7.50 & 7.30 & 0.65 & 0.20 & 0.15 \\
NGC5236-F1-1 & 7.60 & 7.20 & 6.60 & 0.85 & 0.05 & 0.10 \\
\hline
\end{tabular}
\end{minipage}
\end{table}

\begin{figure}
\centering
\includegraphics[width=85mm]{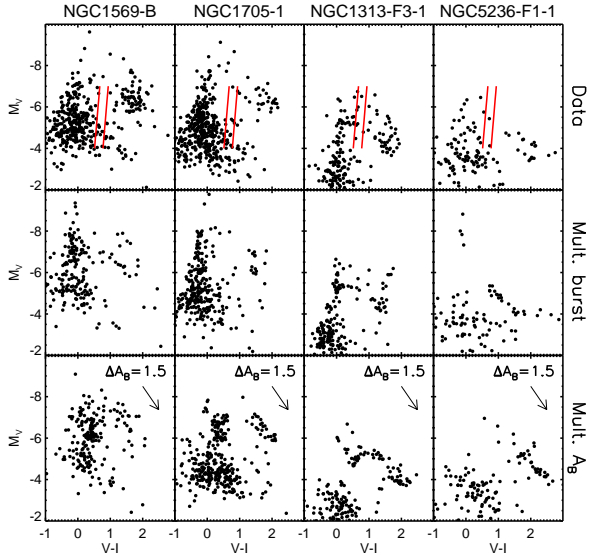}
\caption{\label{fig:mbursts}Top row: Observed CMDs. The red slanted lines indicate the approximate boundaries of the Cepheid instability strip \citep{Sandage2004}. Middle row: CMDs for multiple-burst models (see Table~\ref{tab:mbursts}). Bottom row: CMDs for models with a range of $A_B$ values.}
\end{figure}

\begin{figure}
\centering
\includegraphics[width=85mm]{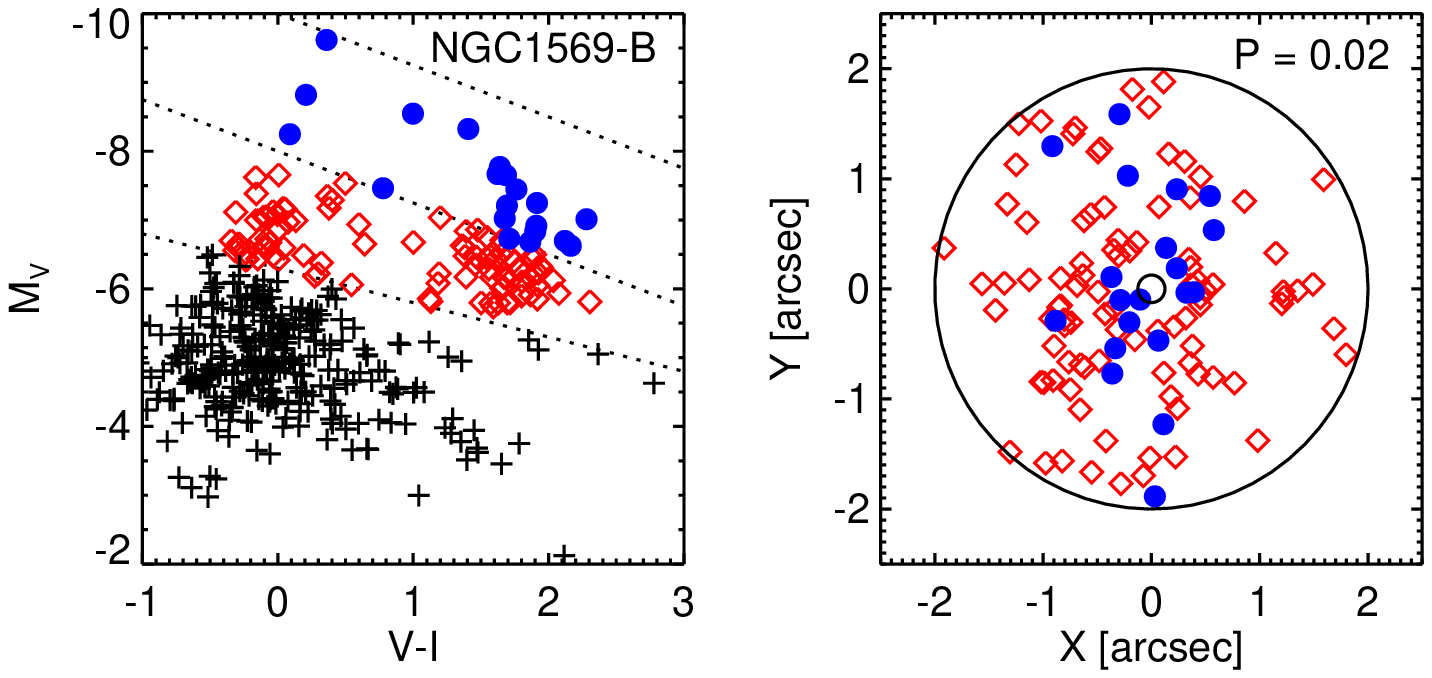}
\includegraphics[width=85mm]{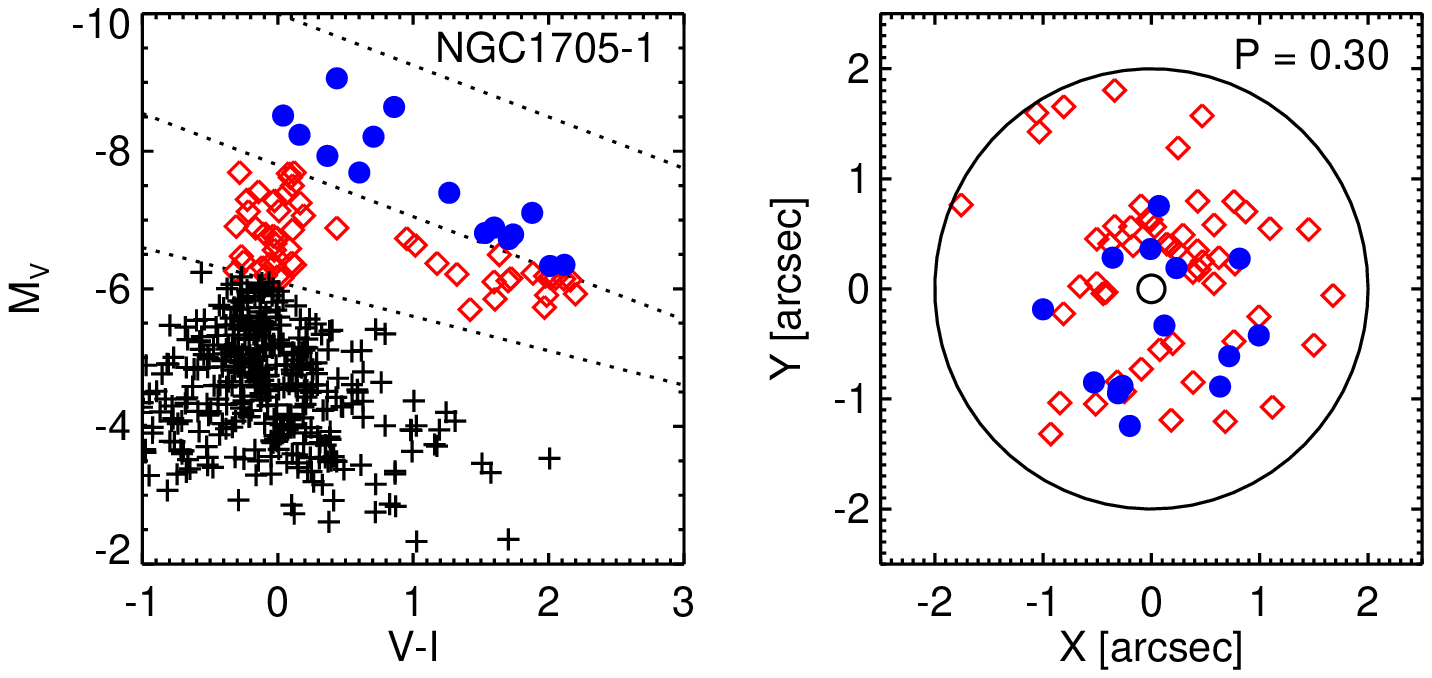}
\includegraphics[width=85mm]{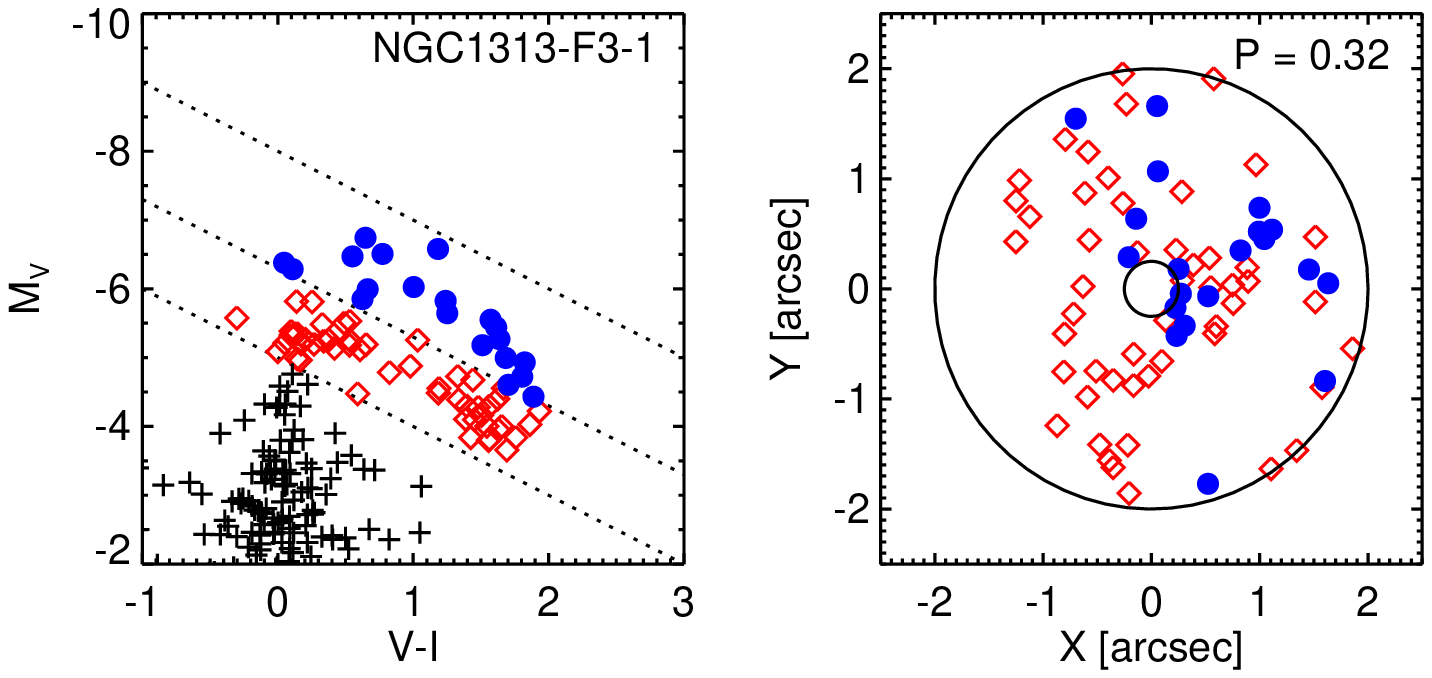}
\includegraphics[width=85mm]{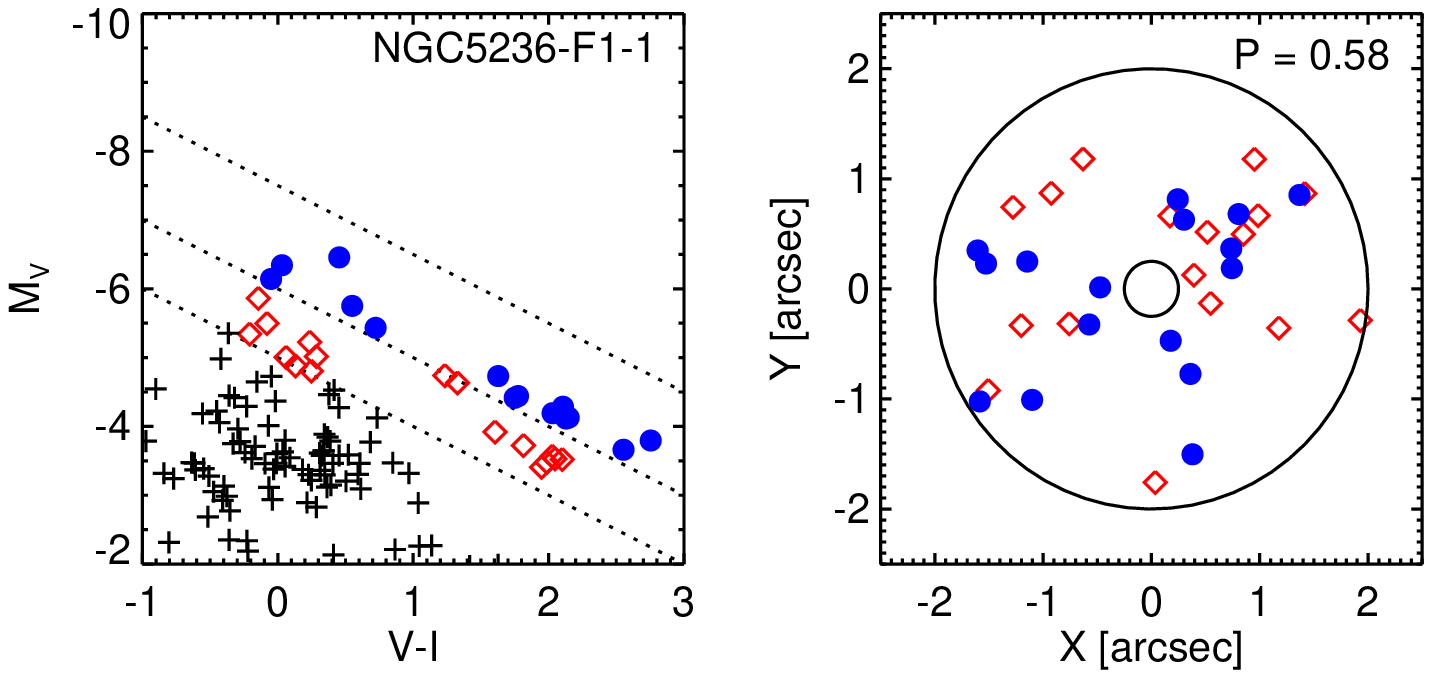}
\caption{\label{fig:spat}Left panels: the division into brighter and fainter supergiant (SG) stars. Right panels: Spatial distribution of the fainter and brighter SGs. K-S tests are generally consistent with no differences in the radial distributions of the two populations, as indicated by the $P$-values in each panel.
}
\end{figure}

In recent years, the traditional view of star clusters as simple stellar populations has been challenged by various pieces of observational evidence. In addition to the long-standing puzzle of anomalous abundance ratios of light elements in globular clusters \citep[e.g.][]{Kraft1979}, this includes multiple main sequence turn-offs and subgiant branches in old GCs \citep{Milone2010}, as well as bi-modal or broad distributions of main sequence turn-off points in intermediate-age clusters in the Magellanic Clouds \citep{Mackey2008,Milone2009} and the presence of dual red clumps in such clusters \citep{Girardi2009}. 
For ancient GCs there are few observational constraints on any actual age spread, and the observed CMD features also appear to require strongly enhanced He abundances \citep{Norris2004,Piotto2005,DAntona2007}.
The LMC and SMC clusters where age spreads have been claimed all have ages of $> 1$ Gyr, and the implied age spreads ($>10^8$ years) would have dramatic effects on the CMDs of younger clusters.
Theoretically, it is a challenge to understand how extended star formation could occur within star clusters on long time scales. After a few Myr, the winds and radiation pressure from OB stars (or eventually, supernovae) are likely to evacuate any remaining gas, and the material subsequently returned by stellar evolution is insufficient to form the observed numbers of second-generation stars unless the first generation has a very top-heavy IMF. This problem may be partly alleviated if a large fraction of the first generation of stars is preferentially lost via dynamical effects, and/or mass lost from stars is supplemented by additional accreted material \citep[e.g.][]{Bekki2011}. It has been proposed that gas may indeed be accreted onto massive clusters as they plough through the interstellar medium of their parent galaxies \citep{Pflamm-Altenburg2009,Conroy2011}.

So far, evidence of age spreads greater than a few Myr is scant in younger clusters. \object{NGC~1850}, one of the most luminous young clusters in the LMC, hosts a small number of stars with ages of only a few Myr that are located within a small subcluster \citep{Fischer1993}. Multiple stellar generations (10--16 Myr and 30--100 Myr) have also been reported for the cluster Sandage 96 in the nearby Sc-type spiral \object{NGC~2403} \citep{Vinko2009} although the older population in this case may be associated with the surrounding field. The giant H{\sc ii} region \object{NGC~604} in \object{M33} appears to host distinct populations with ages of $\sim3$ and $\sim12$ Myr \citep{Eldridge2011}, 
and star formation also seems to have occurred in the star forming region NGC~346 in the Small Magellanic Cloud
over a period of  5--10 Myr \citep{Hennekemper2008,Cignoni2011} and possibly even up to $\sim20$ Myr \citep{DeMarchi2011}. However, the latter two objects are probably too extended to evolve into single bound clusters.

An age spread would result in different main sequence turn-off points within a given cluster, thus providing a way to blur the CMDs, thereby potentially helping to fill the BHG and would also give a spread in the luminosities of the supergiant stars. We therefore now investigate to what extent this would help improve the agreement between  our observations and the synthetic CMDs.

As a starting point, we use the cluster \object{NGC~1313-F3-1}. As can be seen in Fig.~\ref{fig:cmdvi}, the observed colours of the supergiant stars are fairly well approximated by the models for this cluster. It therefore appears plausible that a better overall match to the CMD could be obtained by including a range of ages. We first use the CMD-fitting code FITSFH \citep{Silva-Villa2010a} to search for the combination of isochrones that provides the best match to the observed CMD. This code first constructs a Hess diagram for the observed data, and then attempts to match this by linearly combining Hess diagrams for individual isochrones (such as those in Fig.~\ref{fig:isocmp_p08}) after accounting for data incompleteness and observational errors. 
We used the errors estimated by ALLFRAME as input to FITSFH. Because of the large uncertainties in the completeness estimates, we have not actually included a detailed modelling of detection incompleteness across the CMDs in the SFH fits presented here, but simply restricted the fits to regions of the CMDs above the 50\% completeness limits in Fig.~\ref{fig:cmdvi}.

Fig.~\ref{fig:sfh1313_f3_1} shows the resulting star formation history (SFH) for \object{NGC~1313-F3-1}. The fit suggests a somewhat extended SFH, starting about 50 Myr ago and extending until 15--20 Myr ago. Integrating the SFH in Fig.~\ref{fig:sfh1313_f3_1}, we find that about 2/3 of the stellar mass was formed at $7.6 < \log t < 7.8$, although most of this may have formed in a single burst at $\log t \approx 7.7$. The remaining 1/3 of the mass is in the tail extending to $\log t \approx 7.2$. The first burst coincides quite well with the age derived from the integrated light. 
A similar analysis was done for three other clusters in our sample: \object{NGC~1569-B}, \object{NGC~1705-1} and \object{NGC~5236-F1-1}. For \object{NGC~1569-A} we do not have much freedom to experiment with different SFHs because the lack of RSGs essentially only allows the youngest isochrones, and we further excluded \object{NGC~7793-F1-1} and \object{NGC~5236-F1-3} which have CMDs very similar to those of \object{NGC~1313-F3-1} and \object{NGC~5236-F1-1}, respectively, but with poorer statistics.

While FITSFH will always output some best-fit star formation history, the quality of the ``best fit'' will be limited by the adequacy of the input assumptions, such as the stellar isochrones used, correct assessment of photometric errors, treatment of binaries, etc. 
For young stellar populations, it is generally difficult to obtain a very good fit to observed CMDs \citep[e.g.][]{Silva-Villa2011}, and here we are faced with the additional difficulty of complicated incompleteness effects and relatively small numbers of stars. We cannot, therefore, claim to put very tight constraints on the specific star formation histories of the clusters, but merely aim to explore whether an age spread might account at least in a qualitative way for some of the observed CMD features, and roughly how large it would have to be.
Our final step is therefore to construct model CMDs based on the FITSFH output and compare these with the observed CMDs. As an approximation to the extended SFHs returned by FITSFH, we model each cluster as a superposition of three individual bursts with the ages and relative amounts of mass in each burst listed in Table~\ref{tab:mbursts}. The relative weights of each burst were determined by integrating over the SFHs from FITSFH. We see that, in general, most of the mass is in the oldest burst. The implied age spreads are $\sim10-30$ Myr. Due to the presence of younger populations in each cluster, which decreases the overall mass-to-light ratio of the cluster, the total masses of the multiple-burst models are 25--40\% less than those of the single-burst models listed in Table~\ref{tab:data}.
The modelling as three distinct bursts was done mostly for computational convenience, as it allowed us to easily modify the set-up used for the single-burst models by simply adding three individual ``clusters'' on top of each other. We could, of course, have increased the number of individual ``bursts'', but found that three individual sampling points in age provided an adequately smooth representation in the final CMDs. We emphasize that we do not intend to imply any preference for the clusters actually having experienced distinct bursts over smooth star formation histories.

In Figure~\ref{fig:mbursts} we compare the multiple-burst models with the observations. The top row again shows the observed CMDs, while the middle row shows the CMDs for our multiple-burst models. Compared to the single-burst models, the fits are improved in the sense that we now get a larger range in luminosity for the supergiants, and the BHG is indeed blurred. However, the fits remain less than perfect. For \object{NGC~1569-B}, \object{NGC~1705-1} and \object{NGC~5236-F1-1}, filling the BHG requires the inclusion of a younger isochrone ($\log t_3 = 6.60$) that has no RSGs, so that the number of RSGs is decreased with respect to the single-burst models. At least for \object{NGC~5236-F1-1}, this solution does not appear physically plausible, and results in a number of spuriously bright blue stars in the model CMD.

If the clusters indeed have an age spread, it would be of interest to study the spatial distributions of different sub-populations of stars within them.
In Fig.~\ref{fig:spat} we attempt to make such a comparison, by dividing the supergiant stars into a brighter and fainter subsample in each cluster. The left-hand panels illustrate our adopted division for each of the four clusters, and in the right-hand panels we plot the spatial distributions of the stars. The circles indicate the areas in which photometry was carried out. In \object{NGC~1569-B} there is a hint that the brighter stars appear somewhat more centrally concentrated, but in the other clusters there are no statistically significant differences between the radial distributions of bright and faint stars. This is quantified by the $P$-values from a K-S test, given in each panel.  
We do note that there is a curious asymmetry in the distribution of the brighter stars in \object{NGC~1313-F3-1}. The crossing time is only a few Myr, much less than the age of the cluster, so such substructure should be quickly erased by the internal motions of stars in the cluster. If the brighter stars are as young as $\sim20$ Myr, as suggested by our CMD modelling, and formed as a coherent group, this might help explain how some substructure could remain visible. Alternatively, the apparent asymmetry may simply be the result of a chance alignment.

\subsection{Extinction variations}

In the discussion so far, we have implicitly assumed that there is no significant spread in the extinction towards individual cluster stars. In the bottom row of Fig.~\ref{fig:mbursts} we show simulated CMDs that include a range of $\Delta A_B = 1.5$ mag in the $B$-band extinction. As in the multiple-burst models, this was done by simulating three distinct $A_B$ values ($A_B = 0, 0.75$ and 1.5 mag). In these models, the ages of the model clusters are significantly younger than for the models in Fig.~\ref{fig:cmdvi} and \ref{fig:cmdbv}.  By trial-and-error, we found that the best fits to the observed CMDs were achieved for input ages of $\log t = 7.0, 7.0, 7.50$ and 7.25 for \object{NGC~1569-B}, \object{NGC~1705-1}, \object{NGC~1313-F3-1} and \object{NGC~5236-F1-1}, respectively, and for 70\% of the stars belonging to the most extincted ($A_B=1.5$ mag) bin.

Although a spread in extinction will clearly contribute to blurring of the CMDs, it is evident from the figure that there are several difficulties with this scenario. In all clusters, most of the supergiants would have to be relatively strongly extincted, with only a smaller fraction of them scattering upwards in the CMDs towards lower extinction values. This, however, is in contrast to the main sequence whose colours are well matched by isochrones with little or no additional extinction needed (see also Fig.~\ref{fig:cmdvi} and \ref{fig:cmdbv}). Furthermore, the small width of the main sequence in some clusters (especially \object{NGC~1313-F3-1}, and to some extent \object{NGC~1569-B}), also precludes a large scatter in the extinction values, unless the supergiants are systematically more subject to extinction variations than the main sequence.   
In conclusion, we find it unlikely that extinction variations have a major impact on the observed CMDs.

\subsection{Cepheids}

Another possible reason for the spread in the luminosities of the supergiant stars is that some of these stars are variable. In Fig.~\ref{fig:mbursts} we have indicated the location of the Cepheid instability strip \citep{Sandage2004}. While \object{NGC~1569-B} and \object{NGC~1705-1} are probably too young to host Cepheids, it is possible that some stars in \object{NGC~1313-F3-1} and \object{NGC~5236-F1-1} (as well as \object{NGC~7793-F1-1} and \object{NGC~5236-F1-3}) may be Cepheids. The pulsation amplitudes can amount to more than 1 mag in the $V$-band \citep{Sandage2004}. Although there is a hint that the spread in $M_V$ magnitude is largest near the instability strip, a considerable scatter remains outside the boundaries of the strip, so it is questionable whether this can be the whole explanation. Observations of Cepheids might, however, shed light on possible age spreads within the clusters via the period-age relation \citep{Efremov2003,Bono2005}.

\subsection{Binaries}
\label{sec:binaries}

Binaries are common among massive stars \citep[e.g.][]{Mason2009} and can affect the CMDs in various ways, see for example the excellent discussion by \citet{Pols1994}.  Although it is beyond the scope of this paper to model these effects in detail we will briefly discuss them.

\paragraph{Unresolved multiples:} 
The presence of an unresolved non-interacting companion can increase the brightness by up to a factor two, i.e. change the magnitude by up to $-0.75$ mag.  
 Although we must seriously consider the possibility that all stars have an unresolved companion,  in most cases the companion will be considerably less luminous.  Recent data suggest that the distribution of mass ratios ($q$) is roughly flat, implying that a typical companion is only half the mass of the primary \citep{Pinsonneault2006,Kobulnicky2007,Kiminki2009}. Given the steep dependence of the luminosity on the mass, the effect on the brightness of such a companion would be less than $-0.1$ mag for main sequence stars, and even less if the primary star has already evolved off the main sequence. If unresolved non-interacting binaries were to contribute significantly to any spread in the luminosities of the supergiants, this would require that a significant fraction of these binaries have nearly identical masses so that both stars would be in the supergiant phase at the same time.
 
For these clusters, where the stellar density is high, unresolved higher order multiple systems may be rather common. These may either be physically bound triple or multiple systems or multiple stars that are aligned by chance, although the latter case is in principle taken into account by our artificial cluster experiments. However, even if the light of a star is polluted with the light of two additional equally bright stars the effect on the magnitude is still only  $-1.2$ mag. This is comparable but still smaller than the size of the BHG as predicted by stellar models, which is about $\sim2$ mag.  A shift of 2 mag is reached only for an unresolved multiple system out of which at least 6 stars are equally luminous. We conclude that unresolved multiple systems may contribute to differences between our observed and modelled CMDs
but can certainly not be the lone culprit. 

\begin{figure}
\includegraphics[width=85mm]{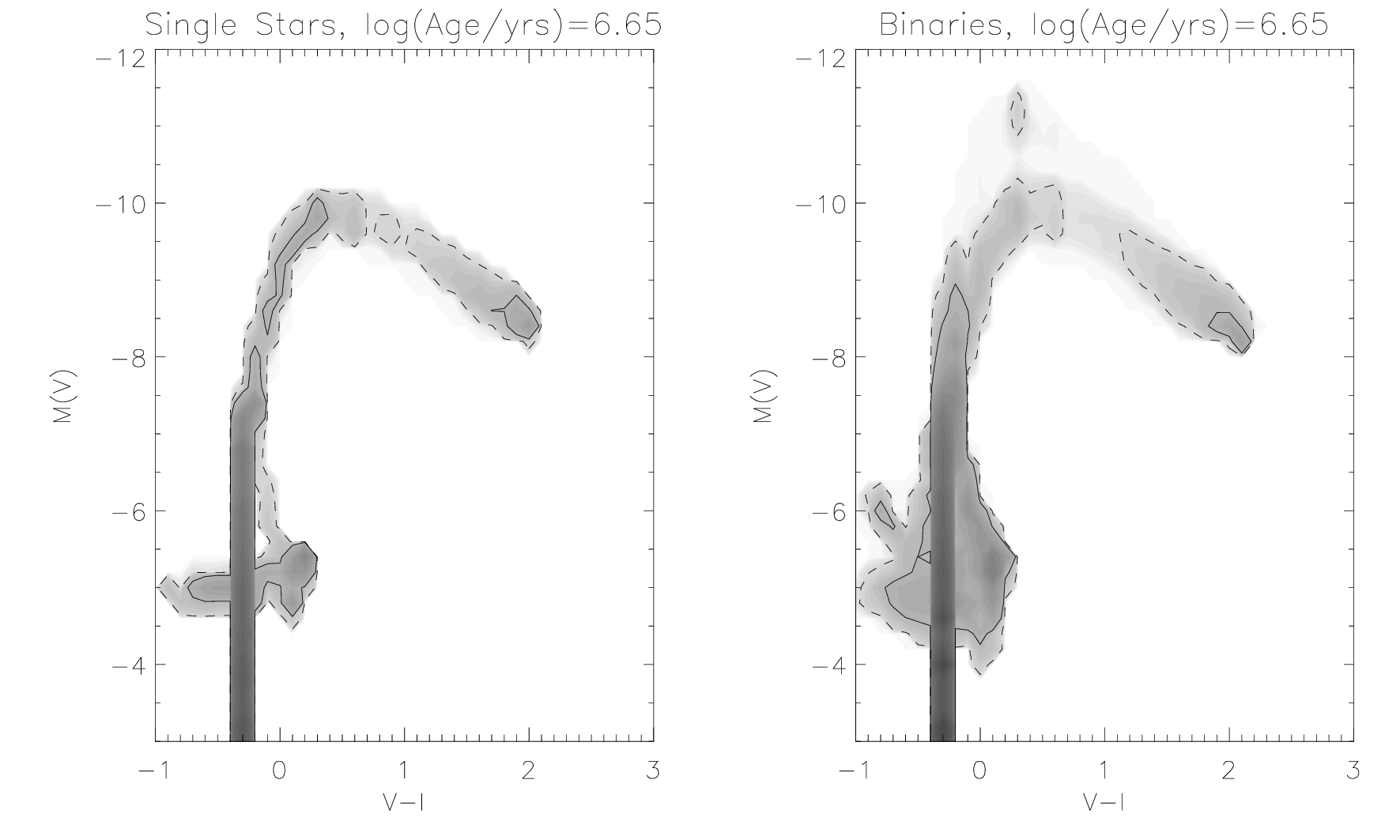}
\includegraphics[width=85mm]{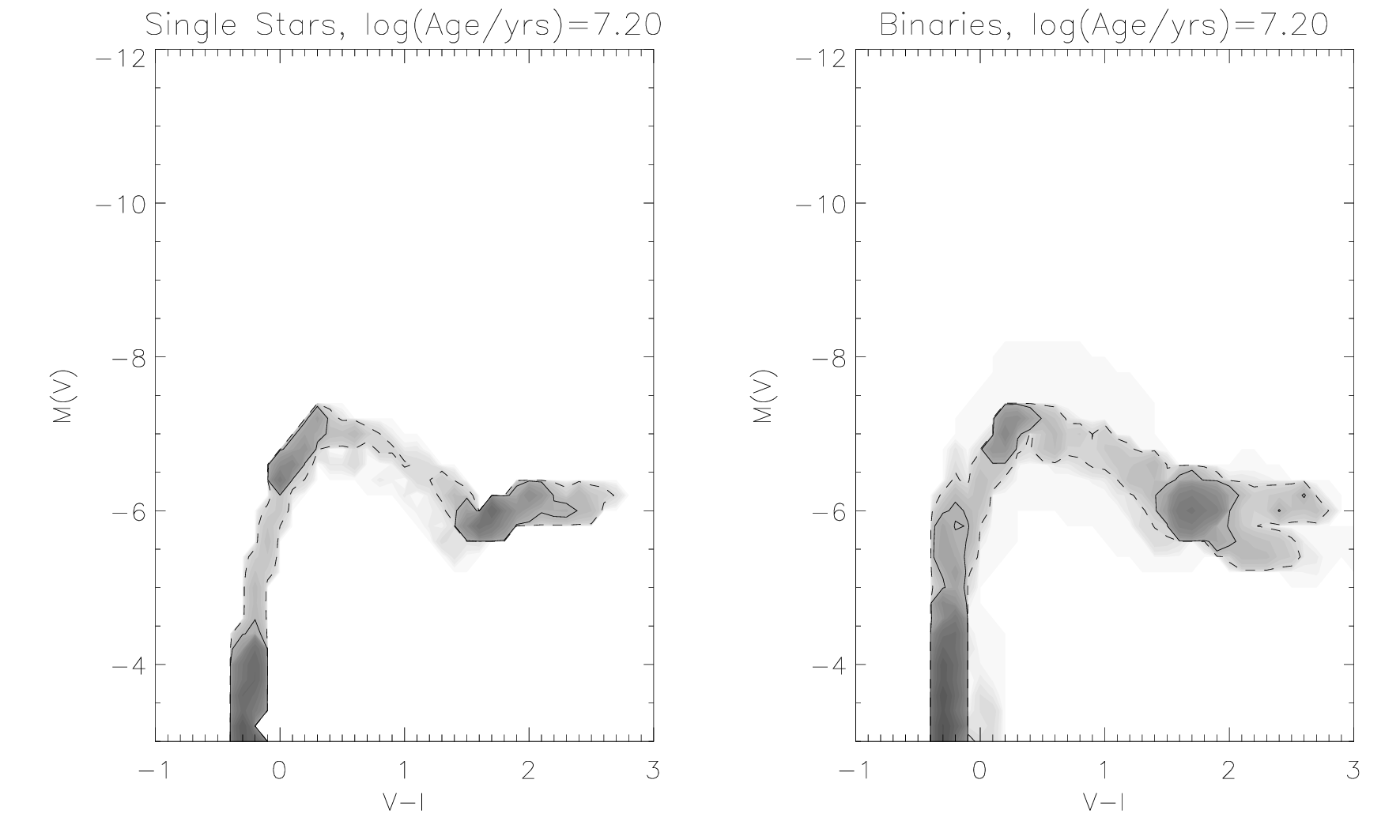}
\includegraphics[width=85mm]{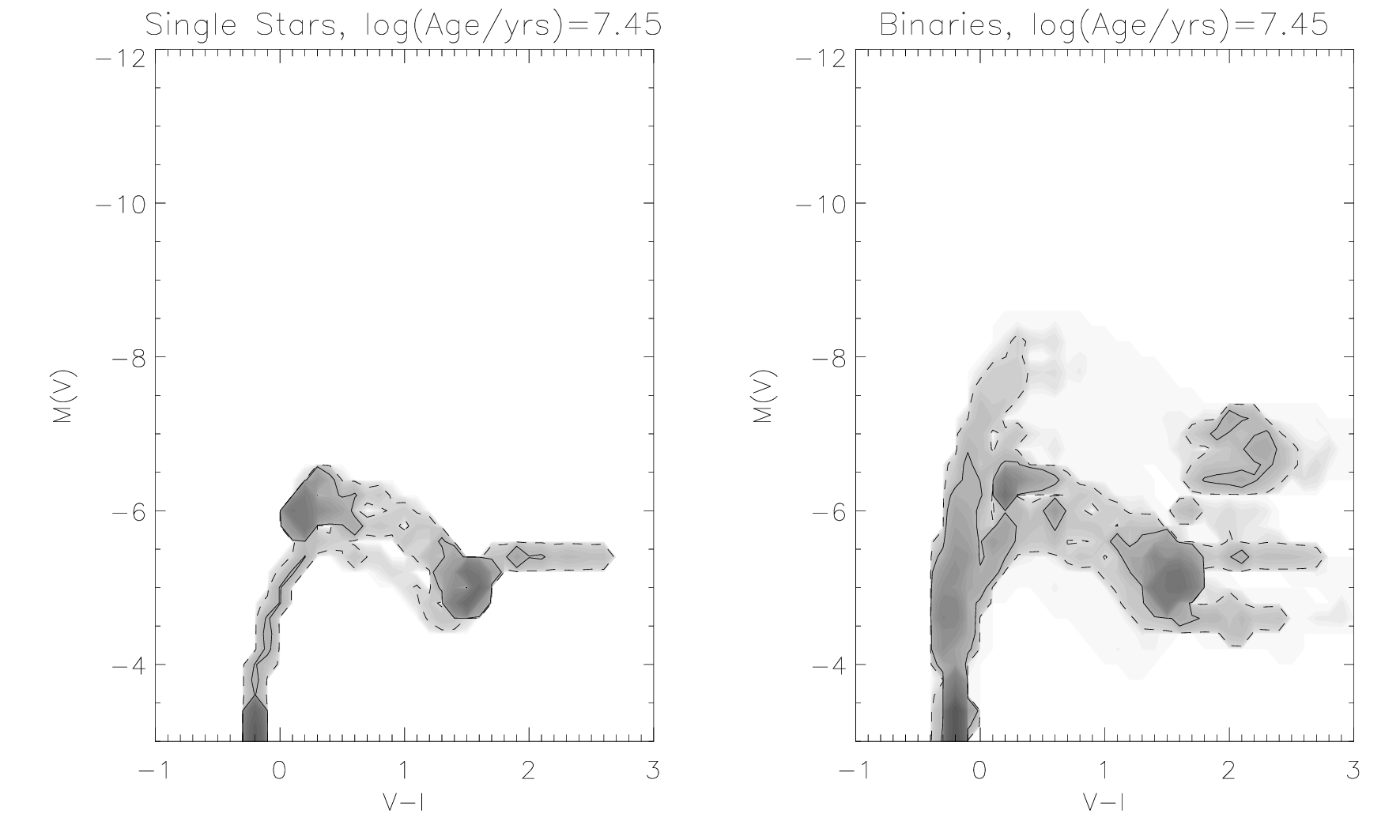}
\caption{\label{fig:toycmd} Example CMDs for single stars (left panels) and binary (right panels) populations calculated with BPASS at three ages typical of clusters in our sample. Due to the nature of the binary models the CMDs are created assuming a small age spread for the population of +/-0.05 dex. The grid cells are 0.1 by 0.2 in $V-I$ and $M_V$ respectively. The grey scale is logarithmic in number of stars with the cell size. The dashed and solid contours represent when the density of stars in each cell is 0.1 and 1 respectively.}
\end{figure}

\paragraph{Interacting binaries:}  
Data from nearby open clusters suggest about 4 out of 10 O stars have companions that are nearby enough to interact \citep{Sana2010}. Even though the fraction of stars whose properties have been altered by binary interaction may not be that large, the consequences of such interactions for the stellar luminosity are potentially severe due to the steep dependence of the luminosity on mass.   In the most extreme cases, interaction can lead to a star that is twice as massive as the turn-off mass, which would imply an increase in the bolometric luminosity of a factor $\sim15$ or $\sim3$ mag. In practice, such cases should be rare and a more typical shift would be on the order of $\sim5\times$, for a star that increases its mass by about 50\% as a result of mass accretion or a merger of the original stars \footnote{assuming that the luminosity of the star after interaction with the mass as $L\sim M^{3.9}$}. The effect on the CMD is still uncertain as it depends on various factors: the distribution of initial binary parameters, the amount of mass loss during interaction or mergers and the life time after interaction.  In order to convert the changes in bolometric luminosity to magnitudes and colours, the effective temperatures of the merger products must also be known.

Binary interaction also affects the post main sequence evolution and therefore the luminosities of red and blue supergiants, as well as their relative numbers.   The presence of a nearby companion can strip the envelope from a star before it reaches giant dimensions, effectively reducing the number of red supergiants by up to 40\% \citep[assuming the distribution functions of][]{Sana2010}.  In addition, the accreting star or merger resulting from binary interaction may under certain circumstances produce blue supergiants as suggested for the progenitor of \object{SN1987A} and the companion of \object{SN1993J} \citep{Hillebrandt1987,Walborn1989,Maund2004}.  This second effect depends, among other things, on the efficiency of internal mixing processes and the number and lifetime of blue supergiants produced is therefore hard to estimate \citep{Claeys2011}. However, in combination the two effects may increase the BSG/RSG ratio by up to a factor of 4 \citep{Eldridge2008}.

In Figure~\ref{fig:toycmd} we show how the CMD might be influenced if approximately two thirds of the stars are in interacting binaries. We use the binary population and spectral synthesis code BPASS \citep{Eldridge2009} to create model stellar populations based on single stars (left) and binaries (right) at three different illustrative ages. 
The models have initial separations between $\log (a/R_{\odot}) = 1$ and 4 in steps of 0.25 dex, while the mass ratio takes values of $q =$ 0.1, 0.3, 0.5, 0.7 and 0.9.
We see that in general the effect of binaries, as discussed above, is to increase the luminosity of stars, including a population of ``blue stragglers'' that populate the region above the MS turn-off, as well as a number of brighter blue and red supergiants. We note that the biggest increase is for the older populations shown here, where it produces an effect very similar to that of an age spread (Fig.~\ref{fig:mbursts}).
These \textit{extra} supergiants arise from binary mergers and secondaries in binaries that become more massive by accreting material from their primaries. In both cases this leads to more massive stars in the stellar population sometime after stars of such a mass would be expected to have exploded in supernovae for a simple single star population. This suggests that the creation of \emph{apparently} younger stellar populations from binary evolutionary paths might plausibly account for some of the differences between our observed CMDs and the models based on canonical stellar isochrones (section~\ref{sec:agespread}). While the code necessarily involves many simplifications in the modelling of binary evolution, and uncertainties in predicting the post-MS evolution of products of binary evolution are at least as large as for single stars, we note that the RSG to BSG ratios for these binary populations in fact do provide values similar to those found in Section \ref{sec:rsgbsg}. However, making more detailed predictions of the effects on the CMDs of these clusters is difficult due to the highly non-linear nature of binary stellar evolution, and beyond the scope of this paper. 

\subsection {Rotational mixing}

Massive stars are observed with a wide range of rotation rates \citep{Penny2009}. Rotation can induce mixing in the stellar interior during core hydrogen burning. This mixing brings fresh fuel towards the core of the star, which leads to an increased life time and to a larger helium core and thus higher luminosity \citep{Maeder2000,Brott2011a}. A distribution of initial rotational velocities will therefore lead to a dispersion of lifetimes and luminosities of stars of the same initial mass.

However, the effect is expected to be rather limited. Even for very rapid rotators, the life time is not increased by more than $\sim 30\%$ \citep{Brott2011b}, and less so for average rotators, which is smaller than the required age spread as discussed in Sect.~\ref{sec:agespread}. The spread in post-main sequence luminosity is predicted to be similar, i.e., 0.3 magnitudes \citep{Maeder2000}, which is again too small to account for the observed luminosity spread in our clusters. Furthermore, while rotational mixing can shift the red edge of the main sequence band slightly redwards, the same is true for the blue edge of the blue loops \citep{Maeder2001}, such that the extent of the predicted blue Hertzsprung gap remains unaffected.

\subsection{The red-to-blue supergiant ratio}
\label{sec:rsgbsg}

\begin{table}
\begin{minipage}[t]{\columnwidth}
\caption{Statistics of blue and red supergiant stars. The ratio $N_{\rm RSG}/N_{\rm BSG}$ is given for the observations, for the single-burst models (SSP) and for the multiple-burst models (Mult.).
}
\label{tab:sgstat}
\renewcommand{\footnoterule}{}
\begin{tabular}{lccccc} \hline\hline
Cluster & $N_{\rm RSG}$ & $N_{\rm BSG}$ & \multicolumn{3}{c}{$N_{\rm RSG}/N_{\rm BSG}$} \\
   & & & Obs. & SSP & Mult. \\ \hline
NGC1569-B & 69 & 45 & $1.53\pm0.09$ & 0.54 & 0.28 \\
NGC1705-1 & 29 & 40 & $0.73\pm0.03$ & 0.40 & 0.16 \\
NGC1313-F3-1 & 39 & 32 & $1.22\pm0.08$ & 1.42 & 1.21 \\
NGC5236-F1-1 & 20 & 13 & $1.54\pm0.30$ & 5.46 & 2.79 \\
\hline
\end{tabular}
\end{minipage}
\end{table}

One long-standing problem in the modelling of massive stars is the number ratio of red to blue supergiants \citep{Langer1995}. Determining this ratio observationally is far from straight-forward. Galactic open clusters generally contain only a small number of such stars, and in the field it is difficult to isolate clean samples of red and blue supergiants because the latter are easily confused with massive main sequence stars. Due to the richly populated post main-sequence phases, the clusters in our sample offer a welcome opportunity to address this issue. 

In Table~\ref{tab:sgstat} we list the numbers of red and blue supergiants ($N_{\rm RSG}$ and $N_{\rm BSG}$) found in the four richest clusters, using the colour division at $(V-I) = 0.8$ (as in Fig.~\ref{fig:lfs}) and the magnitude cuts indicated by the lower dotted lines in Fig.~\ref{fig:spat} (in this comparison we do not distinguish between the ``bright'' and ``faint'' subsamples). Specifically, the cuts are:
\begin{eqnarray*}
  \mbox{\hspace{1cm}NGC~1569-B}: & M_V & < 0.5 (V-I) - 6.3 \\
  \mbox{NGC~1705-1}: & M_V & < 0.5 (V-I) - 6.1 \\
  \mbox{NGC~1313-F3-1}: & M_V & < (V-I) - 5.0 \\
  \mbox{NGC~5236-F1-1}: & M_V & < (V-I) - 5.0
\end{eqnarray*}
Note that the samples of supergiant stars in Table~\ref{tab:sgstat} are larger than those found in any individual Galactic young clusters. We also list the number ratio $N_{\rm RSG}/N_{\rm BSG}$ and its associated error, assuming Poissonian statistics, along with the $N_{\rm RSG}/N_{\rm BSG}$ ratio for our single- and multiple-burst simulations. The random errors on the simulated ratios are very small due to our averaging over 100 realizations of each cluster.
The true uncertainty on the observed $N_{\rm RSG}/N_{\rm BSG}$ is generally larger than indicated by the Poissonian errors, and typically dominated by the difficulty of isolating the BSGs due to the lack of a clear BHG.  For the simulated CMDs in Fig.~\ref{fig:cmdvi}, the adopted cuts make a fairly clean selection of post-MS stars, but for the observed CMDs the number of BSGs depends on where exactly the magnitude cuts are made. 

For \object{NGC~1313-F3-1} the agreement between the observed and simulated $N_{\rm RSG}/N_{\rm BSG}$ ratio is excellent, but for the other clusters there is significant disagreement. For \object{NGC~1569-B} and \object{NGC~1705-B}, the observed ratios are significantly higher than the simulated ones, a discrepancy that is exacerbated in the multiple-age models. For \object{NGC~5236-F1-1} it is very difficult to select a sample of BSGs in a meaningful way.

\subsection{Consequences for integrated colours}

\begin{figure}
\centering
\includegraphics[width=42mm]{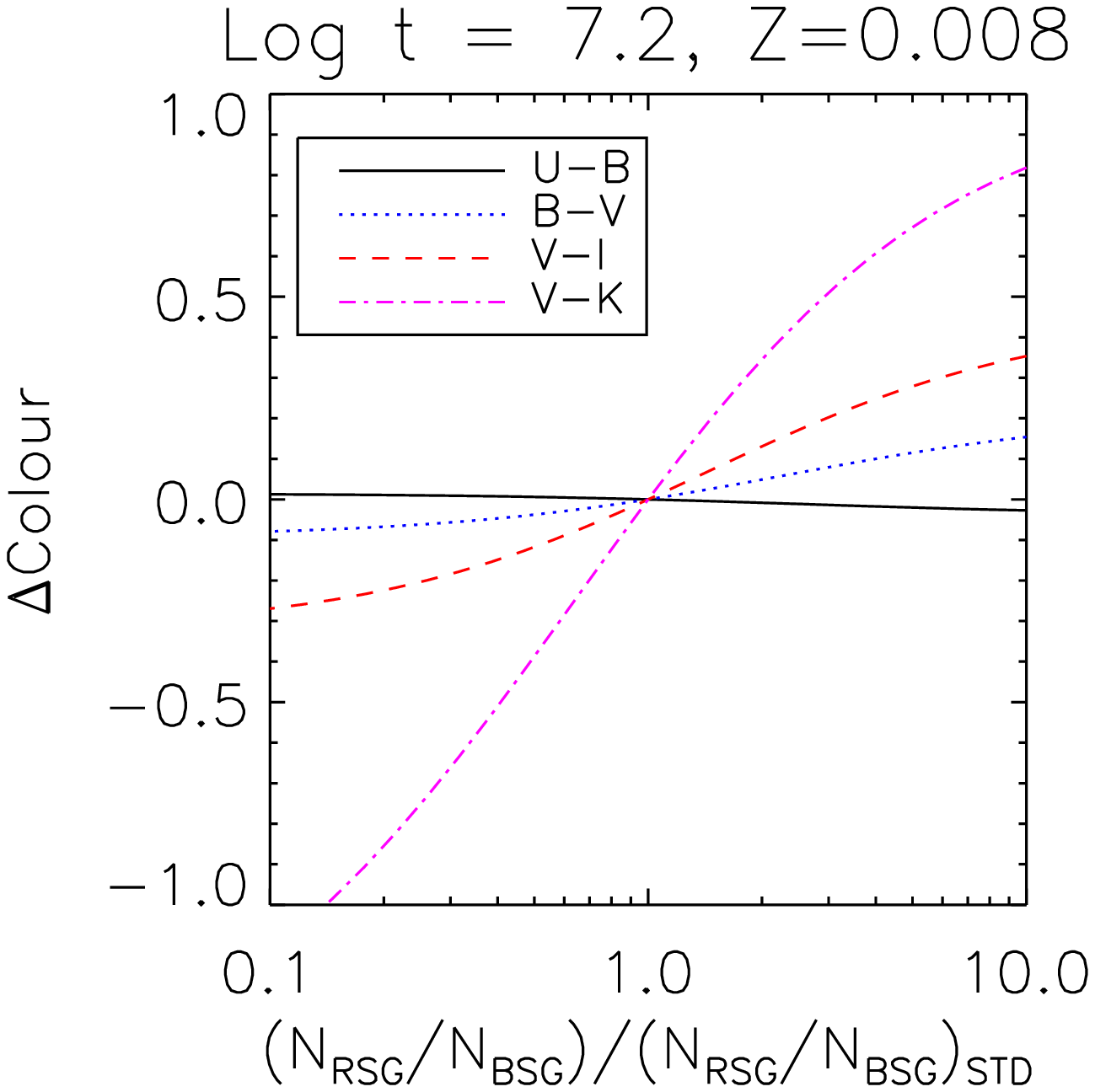}
\includegraphics[width=42mm]{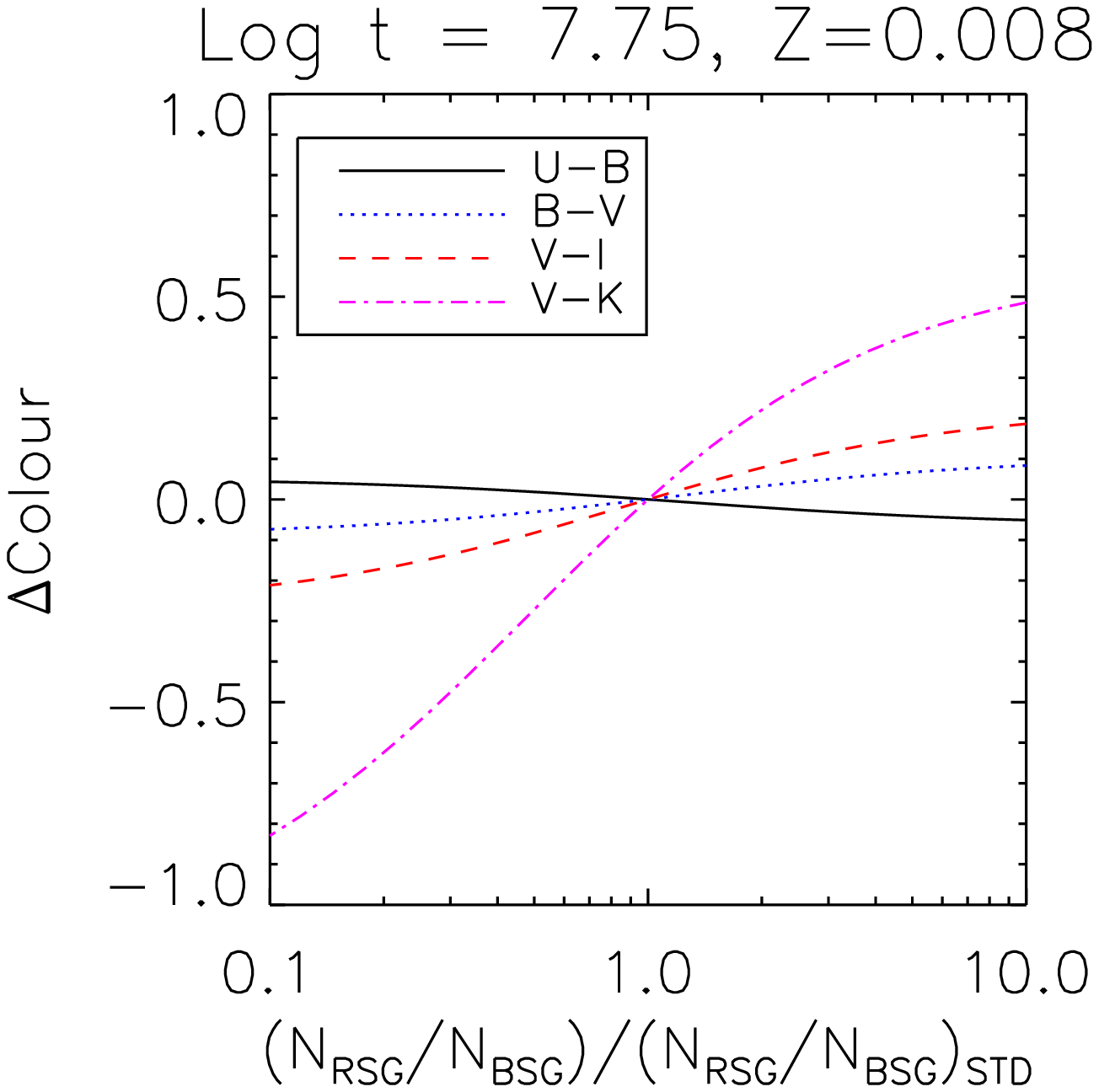}
\caption{\label{fig:rsgbsg}Changes in integrated colours as a function of change in $N_{\rm RSG}/N_{\rm BSG}$ ratio, relative to the ratio predicted by standard isochrones ($N_{\rm RSG}/N_{\rm BSG})_{\rm STD}$.}
\end{figure}

Because the red and blue supergiant stars contribute a significant fraction of the light from a simple stellar population, it is of interest to consider how changes in this ratio will affect integrated properties such as broad-band colours. To assess this, we modified the $N_{\rm RSG}/N_{\rm BSG}$ ratios in the Padua models and re-computed various broad-band colours. We did this by adopting a division between red and blue supergiants at $V-I=0.8$ and changing the relative weights of stars on either side of this limit when computing integrated magnitudes. The (wavelength-dependent) integrated luminosity is then
\begin{equation}
  L_\lambda = \int_{M_{\rm min}}^{M_{\rm max}} w(M) \, L_\lambda(M) \, \xi(M) \, dM
\end{equation}
which is the usual equation for a continuously sampled SSP apart from the factor $w(M)$ which has different values depending on whether a star is on the main sequence or is a red or blue supergiant. Here, $\xi$ is the stellar IMF and $L_\lambda(M)$ is the luminosity of a star with mass $M$. We kept the total number of supergiants constant, i.e.,
\begin{eqnarray*}
 w_{\rm RSG} \, \int_{M \in M_{\rm RSG}}  \xi(M) \, dM +  w_{\rm BSG} \, \int_{M \in M_{\rm BSG}} \xi(M) \, dM \\
  =    \int_{M \in M_{\rm RSG}}  \xi(M) \, dM +  \int_{M \in M_{\rm BSG}}  \xi(M) \, dM
\end{eqnarray*}
Figure~\ref{fig:rsgbsg} shows the result of this exercise for two different ages ($\log t = 7.2$ and $\log t = 7.75$) and four different broad-band colours ($U-B$, $B-V$, $V-I$ and $V-K$). This calculation was done for actual Johnson-Cousins colours. We plot the change in each colour with respect to the colour computed for the unmodified isochrones, as  a function of the change in the $N_{\rm RSG}/N_{\rm BSG}$ ratio (i.e., $w_{\rm RSG}/w_{\rm BSG}$). For $B-V$, $V-I$ and $V-K$ we see that the integrated colours indeed do become redder when the $N_{\rm RSG}/N_{\rm BSG}$ ratio is increased. However, for $U-B$ the opposite happens. This somewhat counterintuitive result may be understood as follows: Taking the standard $\log t = 7.75$ model as an example, the contribution of the RSGs to the integrated light is (0.8\%, 4.0\%, 13\%, 33\%, 70\%) in the $(U, B, V, I, K)$ passbands, while the BSGs contribute (17\%, 25\%, 26\%, 23\%, 13\%) in the same bands. 
Because both the blue and red supergiants are cooler than the main sequence stars, the net effect of both is to produce redder integrated colours than would be the case for a cluster without any supergiant stars. At long wavelengths the red supergiants dominate the light, and increasing the $N_{\rm RSG}/N_{\rm BSG}$ therefore has the expected effect of making the integrated colours redder. However, at short wavelengths the contribution from the RSGs to the integrated light becomes very small, and only BSGs are able to produce a significant change in the integrated colours. Hence, increasing the $N_{\rm RSG}/N_{\rm BSG}$ ratio in this case has the net effect of decreasing the overall contribution of supergiant stars to the integrated light at shorter wavelengths, thus making the integrated $U-B$ colours bluer rather than redder.

Fortunately, the changes in integrated photometry appear relatively modest even for rather extreme (factor of 10) changes in the $N_{\rm RSG}/N_{\rm BSG}$ ratio, at least for optical colours. $U-B$ is the most age-sensitive of the colours shown in Fig.~\ref{fig:rsgbsg} (at least at young ages), and a change of $\Delta (U-B)$ of 0.1 mag corresponds to a difference of about 0.1 dex in the age of a cluster. It therefore appears that age estimates from integrated colours may not be too strongly affected by uncertainties in the RSG/BSG ratio. However, it would be worth investigating in more detail how other properties estimated from integrated colours (e.g. extinction, metallicities, etc.) would be affected. In particular, the large effect on the near-IR colours might lead to substantial uncertainties in fits involving these colours that attempt to simultaneously constrain age, metallicity and extinction. Combined with the fact that
age determinations involving near-IR colours are also more sensitive to stochastic sampling of the stellar IMF
\citep{Fouesneau2010}, this implies that the inclusion of more passbands (in particular extending to the near-IR) does not necessarily lead to a better determination of physical cluster properties from integrated colours.

\section{Summary}

We have presented resolved photometry for seven young massive star clusters in five nearby galaxies, located at distances of 3--5 Mpc. While it is challenging to carry out photometry for individual stars in clusters at such distances, the richness of these clusters allows us to obtain colour-magnitude diagrams for large samples of supergiant stars. In order to assess the effects of crowding, we have carried out extensive artificial-cluster tests where we generated synthetic clusters with properties as similar as possible to those of our real clusters. By comparing the colour-magnitude diagrams of our real and simulated clusters, we reach the following conclusions:

\begin{itemize}
\item We confirm previously known problems with canonical stellar isochrones:
\begin{enumerate}
  \item The standard models predict a clear separation in the CMD between the H-burning main sequence and the He-burning ``blue loop'' stars (i.e., the blue Hertzsprung gap) while no such gap is seen in the data. 
  \item The observed red-to-blue supergiant ratio is in general not well reproduced by the models. However, the lack of a clear BHG makes it difficult to identify clean samples of blue supergiants for this comparison
\end{enumerate}
\item We have investigated the effect of changing the red-to-blue supergiant ratio on integrated colours and find that even large changes in this ratio have relatively modest effects on optical colours that would affect age estimates from integrated photometry by less than 0.1--0.15 dex. However, colours involving infrared passbands (such as $V-K$) are much more strongly affected, strengthening the conclusion that inclusion of such colours may actually increase, rather than decrease, the systematic errors when deriving cluster properties from integrated photometry \citep{Fouesneau2010}.
\item We observe a significant scatter ($\sim2$ mag) in the luminosities of the supergiant stars in some clusters. Comparison with our simulated clusters indicates that this scatter cannot be explained by observational errors alone. 
It may be partly due to variability, since some of the clusters are in the age range where Cepheid variables are expected. However, the scatter is seen also for stars outside the instability strip.
\item By including an age spread of 10--30 Myr in our model clusters we can achieve a better match to the observations. This is, however, much smaller than the spread that has been reported for older clusters in the Magellanic Clouds, and an age spread cannot fully explain the morphologies of all the CMDs. 
While uncertainties in stellar models may account for the differences between predictions and observations of the BHG, this is a less likely explanation for a spread in the supergiant luminosities.
We discuss several other mechanisms that might contribute to differences between observed and model CMDs, and conclude that extinction variations and unresolved binaries are unlikely to be the main culprits, while interacting binaries might also have a significant effect on the CMDs. 
\item In general, we find no significant differences in the spatial distributions of the brighter and fainter supergiant stars. In \object{NGC~1569-B} there is a hint that the brighter stars are somewhat more centrally concentrated, and in \object{NGC~1313-F3-1} there is a curious asymmetry in the distribution of the brighter stars.
\end{itemize}

Each of the clusters studied in this paper has its own unique characteristics and the current sample of seven objects is clearly still too small to draw general conclusions. However, our search of the HST archive is far from complete, and it is likely that several additional objects have already been observed that would lend themselves to similar analysis. Furthermore, the ongoing Panchromatic Hubble Andromeda Treasury survey \citep{Dalcanton2011} will include a vast number of young clusters that may also be useful for addressing the issues discussed here (although most of them significantly less massive than those included here).
Finally, it is our hope that the data presented here will prove useful as benchmarks for developers of stellar models.

\begin{acknowledgements}
We thank Livia Origlia for discussion and comments, and the anonymous referee for a careful reading of the manuscript and several suggestions that helped us clarify the presentation. SdM is supported by NASA through Hubble Fellowship grant HST-HF-51270.01-A awarded by the Space Telescope Science Institute, which is operated by the Association of Universities for Research in Astronomy, Inc., for NASA, under contract NAS 5-26555. SSL acknowledges support from the Nederlandse Organisatie voor Wetenschappelijk Onderzoek (NWO) via a VIDI grant.
\end{acknowledgements}

\bibliographystyle{aa}
\bibliography{libmen.bib}

\end{document}